\documentclass[11pt]{article}

\usepackage[english]{babel}
\usepackage[margin=2.5cm]{geometry}
\usepackage{setspace}
\usepackage{graphicx}
\usepackage{color}
\usepackage[breaklinks]{hyperref}
\usepackage[defaultlines=4,all]{nowidow}
\usepackage{tabularx}
\usepackage{soul}
\usepackage{amsmath}
\usepackage{dcolumn}
\usepackage{array}
\usepackage{lscape}
\usepackage{geometry}
\usepackage{csquotes}

\usepackage[backend=biber,citestyle=authoryear,isbn=false,url=false,eprint=false,sorting=none,uniquename=init]{biblatex}
\title{Universal role of commuting in the reduction of social assortativity in cities}
\author{Eszter Bokányi$^{1,2,*}$, Sándor Juhász$^{1,2}$, Márton Karsai$^{3,4}$, Balázs Lengyel$^{1,2}$}

\date{{\footnotesize%
    $^1$Corvinus University of Budapest; Laboratory for Networks, Technology and Innovation\\%
    Budapest, H-1093, Hungary\\%
    $^2$ELKH Centre for Economic and Regional Studies, Agglomeration and Social Networks Lendület Research Group\\%
    Budapest, H-1097, Hungary\\%
    $^3$Central European University; Department of Network and Data Science\\%
    Vienna, A-1100, Austria\\%
    $^4$Rényi Alfréd Institute of Mathematics,\\%
    Budapest, H-1053, Hungary\\%
    $^*$Corresponding author: \href{mailto:bokanyi.eszter@krtk.hu}{bokanyi.eszter@krtk.hu}}\\[2ex]%
}

\begin{document}

\maketitle

\onehalfspacing

\begin{abstract}
\noindent
Millions commute to work every day in cities and interact with colleagues, customers, providers, friends, and strangers. Commuting facilitates the mixing of people from distant and diverse neighborhoods, but whether this has an imprint on social inclusion or instead, connections remain assortative is less explored.
In this paper, we aim to better understand income sorting in social networks inside cities and investigate how commuting distance conditions the online social ties of Twitter users in the 50 largest metropolitan areas of the United States. Home and work locations are identified from geolocated tweets that enable us to infer the socio-economic status of individuals. Our results show that an above-median commuting distance in cities is associated with more diverse individual networks in terms of connected peers and their income. The degree that distant commutes link neighborhoods of different socio-economic backgrounds greatly varies by city size and structure. However, we find that above-median commutes are associated with a nearly uniform, moderate reduction of social tie assortativity across the top 50 US cities suggesting a universal role of commuting in integrating disparate social networks in cities. Our results inform policy that facilitating access across distant neighborhoods can advance the social inclusion of low-income groups.
\end{abstract}


\section{Introduction}

Cities are champions of diversity \parencite{jacobs2016death, glaeser2011cities, bettencourt2013origins}. Complex interaction networks of individuals in urban areas enabled by population density, co-location, and easy access together made cities the global engines of technological and economic progress \parencite{duranton2020economics,storper2004buzz,calabrese2011interplay, chong2020economic}. However, cities are also known for high levels of segregation \parencite{Sampson2008, glaeser2009inequality, FloridaMellander2015} where disparate neighborhoods are separated from each other in the urban space \parencite{ananat2011wrong, Chodrow2017, FryTaylor2012, Bokanyi2016, Massey1988}. Furthermore, spatial segregation by income also fragments social networks, which can hinder progress and can deepen inequalities \parencite{Eagle2009, Bailey2020, Norbutas2018, abitbol2020interpretable, toth2021inequality}. Given the importance of this problem, a growing community has investigated the patterns of mobility in cities to better understand mixing potentials across disparate and diverse neighborhoods \parencite{Wang2018, pappalardo2015using, Dong2019, Heine2021},
which may increase economic prosperity \parencite{eagle2010network}. Yet, less is known whether mobility mixing has any imprint on the social connections of people.

Commuting covers a large share of urban mobility \parencite{jiang2016timegeo} and by connecting home with work locations, the places where people spend most of their time, it plays an important role in the spatial formation of social connections \parencite{Dahlin2008, Calabrese2011, Small2019}. Since aggregated social networks form spatially bounded communities across neighborhoods \parencite{Bailey2020}, the further one commutes, the higher the likelihood that commuting-related social connections will introduce diversity in the egocentric network of the commuter \parencite{Viry2012, Blumenstock2019}. Due to spatial segregation, economically disparate neighborhoods tend to be far from each other \parencite{roberto2015spatial}, thus long commutes are more likely to link places with different social status \parencite{Ham2018, Nieuwenhuis2020}. Nevertheless, it is not trivial that long commutes should facilitate social inclusion, because social interactions might remain assortative even at places far from home \parencite{Wang2018, Morales2019, Dong2019}. Meanwhile, the time to develop new social connections is especially limited for low-income workers who travel to work during rush hours \parencite{florez2018measuring, Dannemann2018}.

The spatial distribution of high versus low-income households determines the length of travel that can bridge disparate neighborhoods. Since the scale of socio-economic isolation greatly varies across cities \parencite{Chodrow2017}, one may expect that the mobility of people also enables a different degree of social mixture. However, the assortativity of urban mobility is a universal feature across cities: individuals have been recently reported to visit locations that are similar to their home neighborhood \parencite{Bora2014, Wang2018, Dong2019, Leo2016, Yip2016}. Yet, how assortativity of commuting and social networks are related and how this relation is modified by the length of commute in cities is still largely uncovered. 

In this paper, we aim to better understand how mixing in urban social networks is facilitated by commuting. To answer this question, we use a unique dataset on 348,850 Twitter users living in the 50 largest metropolitan areas of the US and track their home and work locations as well as their mutual followership ties on the platform, which from now on, we call the social network of users. We project these social networks in the urban space and attribute users with an average income based on their home locations on an income map extracted from census data. By comparing ego network indicators between people commuting to different distances, we find that long commuting is associated with lower levels of transitivity, the tendency that friends of friends know each other, and higher levels of income diversity among friends. These results are consistent across the 50 largest US cities and suggest that long commutes can indeed facilitate social mixing.

Our results suggest a universal role of commuting in integrating disparate social networks. The paper contributes to the discussion on the importance of commuting in cities and shows that longer commutes have a measurable even though moderate influence on establishing diverse and less segregated social connections. The findings imply that supporting access to distant work can help the inclusion of lowest income groups and to a certain degree the richest as well, regardless of the urban context.

\section{Results}

We use a unique Twitter database that contains all messages and profile information of 348,850 Twitter users in the top 50 metropolitan areas of the United States. The data was collected between 2012 and 2015 and due to the sample selection method described in \cite{Dobos2013}, the database contains a considerable amount of individuals who allowed automatic GPS data collection for all their messages. This dataset was used in previous research to detect dominant language use and temporal patterns connected to socio-economic indicators such as ethnicity or unemployment in the US, to establish world-wide communities of users reflecting political and cultural boundaries, and to model the spreading of viral content \parencite{Bokanyi2016, Kallus2015, Kallus2017, bokanyi2017prediction}. 

\begin{figure}[h!]
\centering
\includegraphics[width=\linewidth]{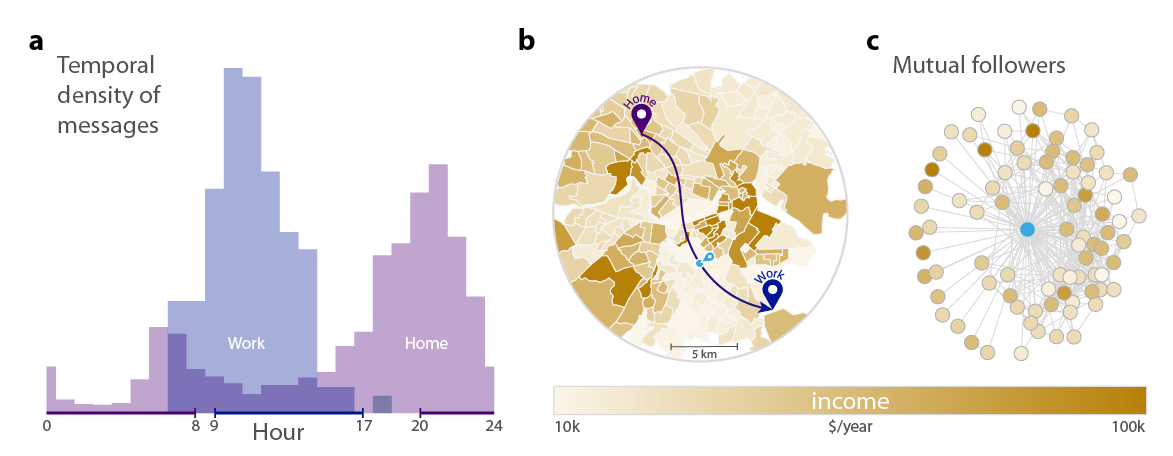}
\caption{\footnotesize 
Combination of spatial, temporal and social network data of geolocated Twitter messages. (a) Home and work locations of users are identified through the distribution of timestamps on all their collected tweets within their most frequently visited spatial clusters. We assign a possible home location (8PM-8AM) and a possible work location (9AM-5PM) to each user \parencite{Lambiotte2008, McNeill2017} as their most frequently visited location in the given period. The histogram represents the timeline of tweets for the clusters of a sample user.
(b) Commuting is defined as the overhead distance between users' home and work locations. The colorbar of the map indicates the income level of census tracts.
(c) Twitter ego network of a sample user based on mutual followership. The coloring of nodes also corresponds to the level of income in the home tract of users.
}
\label{fig:fig1}
\end{figure}

Figure \ref{fig:fig1} illustrates how commuting and social network information is retrieved from the data. Home and work locations are detected by the most frequent locations of tweets in the morning and evening hours or during daytime as depicted in Figure~\ref{fig:fig1}a (and as explained in Materials and Methods). This process enables us to identify the census tract of home and work locations and attach socio-economic status, measured by the average household income of census tracts from the 2012 American Community Survey. Commuting is characterized by the Euclidean distance between home and work and the socio-economic status of both locations. Finally, we construct the ego network for every user from mutual followership of Twitter profiles and characterize egos and alters by the socio-economic status of their home location. This enables us to quantify social mixing in terms of commuting and social ties in cities. 

Figure~\ref{fig:fig1}b shows the census tracts of inner Boston colored by the average annual household incomes and the home and work locations of a sample user. The user's ego network is depicted in panel (c), with colors indicating the income of the neighbors inferred from their home census tract. Each user in our sample has at least 1 mutual followership-based connection and has identifiable home and work locations that are at least 100~metres away from each other. The distribution of users across the 50 selected cities is illustrated in Supplementary Information (SI) 1 and 2. For a more detailed description, see Materials and Methods.

To characterize the relation between $d$, the distance of commutes, and the social network of individuals, we compare the social networks of people commuting to $d>{median}$ with $d<{median}$ commuting distances in each of the 50 largest US metropolitan areas. Median commuting distances are calculated on the basis of the sampled users in each city as illustrated in SI 3. Our expectation is that commuting may induce more out-of-community independent social ties for commuters, in turn decreasing the transitivity of their egocentric networks. We observe this effect by measuring the local clustering coefficient \parencite{Watts1998} for each user, which quantifies the tendency that an individuals' friends know each other. Another assumption of ours is that these out-of-community ties introduce stronger diversity in ego networks in terms of socioeconomic status of neighbors. We quantify this effect via the average income difference from friends in users' ego networks, which measures the income similarity of online social connections (for a formal definition see Eq.~\ref{eq:income} in Materials and Methods).

\begin{figure}[ht!]
\centering
\includegraphics[width=\linewidth]{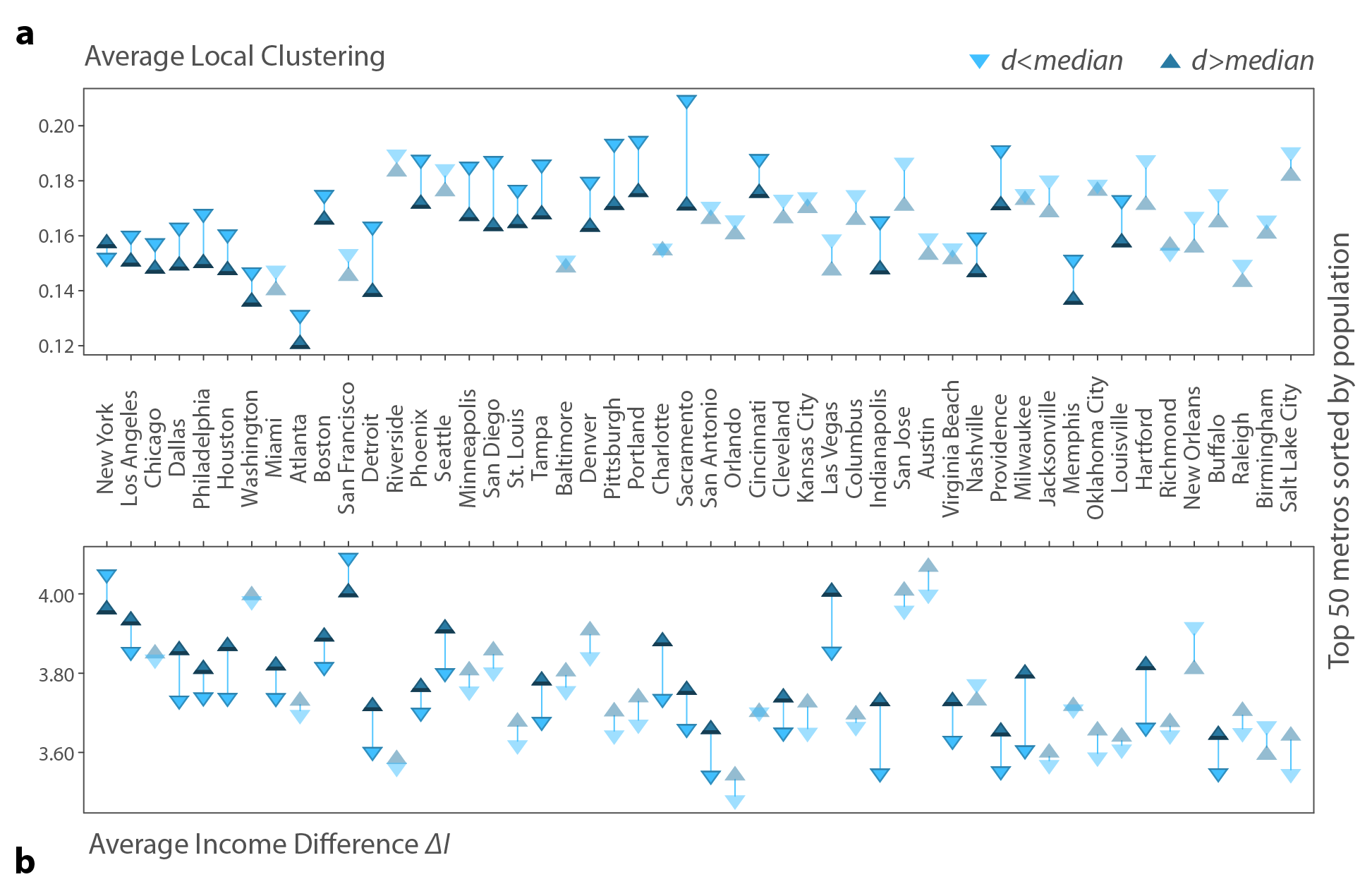}
\caption{\footnotesize Network characteristics of users and commuting distance in the top 50 metropolitan areas of the United States. (a) Network closure measured by the local clustering coefficient is lower in most cities for those users who commute further than the local median distance. (b) Income mixture, measured by average income difference from friends, is higher of those who commute above the local median distance in the majority of metropolitan areas. Non-transparent symbols indicate that the differences of means are significant (p$<$0.05).}
\label{fig:fig2}
\end{figure}

Figure~\ref{fig:fig2}a reports the average of local clustering coefficient and (b) the average income differences of users commuting above and below the local median distance in the 50 largest metropolitan areas in the USA. These findings suggest that, with a few exceptions, an above-median distance commute is associated with lower local clustering (Figure~\ref{fig:fig2}a), and with greater income difference in the commuters' ego networks (Figure~\ref{fig:fig2}b). This implies that working further away from home helps people to develop less cohesive and income-wise more diverse social networks in most metro areas. Note that here metropolitan areas are sorted in decreasing population order and non-transparent markers denote significant differences (p $<0.05$) between averages. 

While these results suggest clear trends, they also highlight the heterogeneity of cities. To support these observations, in SI 4, we compute the degrees for below and above median distance commuters, and we also repeat the measurements and find them to be robust for various distance thresholds. A multivariate regression analysis using continuous variables in SI 5 provides further evidence that commuting distance correlates negatively with local clustering even when controlling for the number of connections and income. These regressions also inform us that commuting distance facilitates mixing in social networks by enabling commuters to make more friendships.

\begin{figure}[b!]
\centering
\includegraphics[width=\linewidth]{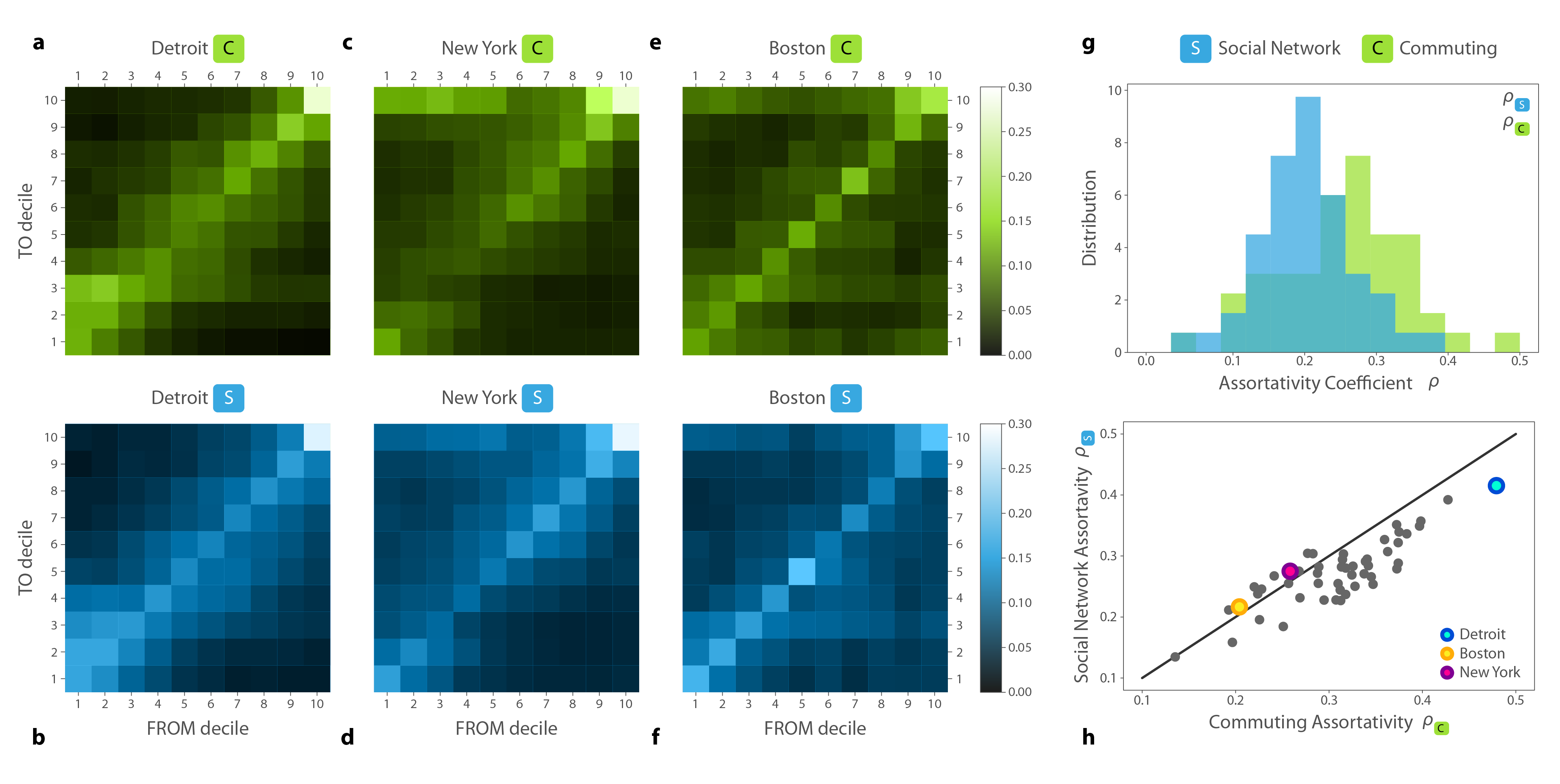}
\caption{\footnotesize (a) Commuting assortativity matrix $C$ and (b) social network assortativity matrix $S$ between the 10 income deciles for Detroit, New York (c) and (d) and Boston (e) and (f). (h) Distribution of Pearson correlations $\rho_C$ (green) and $\rho_S$ (blue) for the assortativity matrices $C$ and $S$ of the top 50 metropolitan areas of the US. (g) Commuting assortativity and social network assortativity are strongly correlated across cities. Solid line represent $\rho_C=\rho_S$.}
\label{fig:fig3}
\end{figure}

For a more detailed insight into the structure of social and mobility assortativity in these cities, next, we analyze social mixing through commuting and online social ties between income groups. We sort all census tracts into income deciles based on the income distribution across all census tracts in the metro area in question and assign an income decile ranging from 1 to 10 to home and work locations. For each metro area, we construct a commuting assortativity matrix $C$ and a social network assortativity matrix $S$ to represent connection probabilities between these income deciles. The elements of the commuting assortativity matrix $C_{ij}$ measure the probability that a user with home census tract in income decile $i$ commutes to work in a census tract of income decile $j$. Similarly, elements of the social network assortativity matrix $S_{ij}$ represent the average probability that a person living in a tract with income decile $i$ has a mutual followership tie with a user living in a tract with income decile $j$. For more details on the construction of the matrices, see Materials and Methods.

The aggregated patterns of commuting $C$ and friendship ties $S$ are presented in Figures~\ref{fig:fig3}a-f for three example metropolitan areas, Detroit, New York, and Boston. Unlike previous studies \parencite{Dong2019, Morales2019}, we do not observe universal assortativity patterns over all cities in these networks.
In some of the cities, such as Detroit, the strong diagonal component features strong segregation patterns, meaning that people tend to commute to neighborhoods with similar annual household incomes as their home neighborhood, and they tend to form social ties with people living in neighborhoods with similar income, as also found in \parencite{Heine2021}. In cities like Boston, patterns of mobility and online social ties are less assortative with higher likelihood for diverse, off-diagonal connections. All commuting and social network matrices are available in the SI 6 for the 50 metropolitan areas. 

To explore this heterogeneity further, we computed the Pearson correlation coefficient of the above matrices (see Materials and methods equation \eqref{eq:rho}).  We use these correlation coefficients as a single-number measure of assortativity in the metropolitan-level networks  denoted by $\rho_C$ for the commuting, and $\rho_S$ for the social network assortativity matrix. We show the $\rho_C$ and $\rho_S$ distributions in Figure~\ref{fig:fig3}g. We see here that the level of assortativity varies remarkably across the 50 metro areas, but judging by their averages, commuting in metro areas ($\bar{\rho}_C=0.31\pm0.07$) are more income assortative than online social ties ($\bar{\rho}_S=0.27\pm0.05$). Interestingly, our observations in Figure~\ref{fig:fig3}a-f further suggest that the measured commuting and social network assortativity matrices are not independent of each other. Indeed, Figure~\ref{fig:fig3}g illustrates that $\rho_C$ and $\rho_S$ pairs are strongly correlated ($\rho$=0.84) suggesting a substantial relationship that social networks are segregated in cities where home-work commuting patterns are assortative. 

\begin{figure}[htbp!]
\centering
\includegraphics[width=0.9\linewidth]{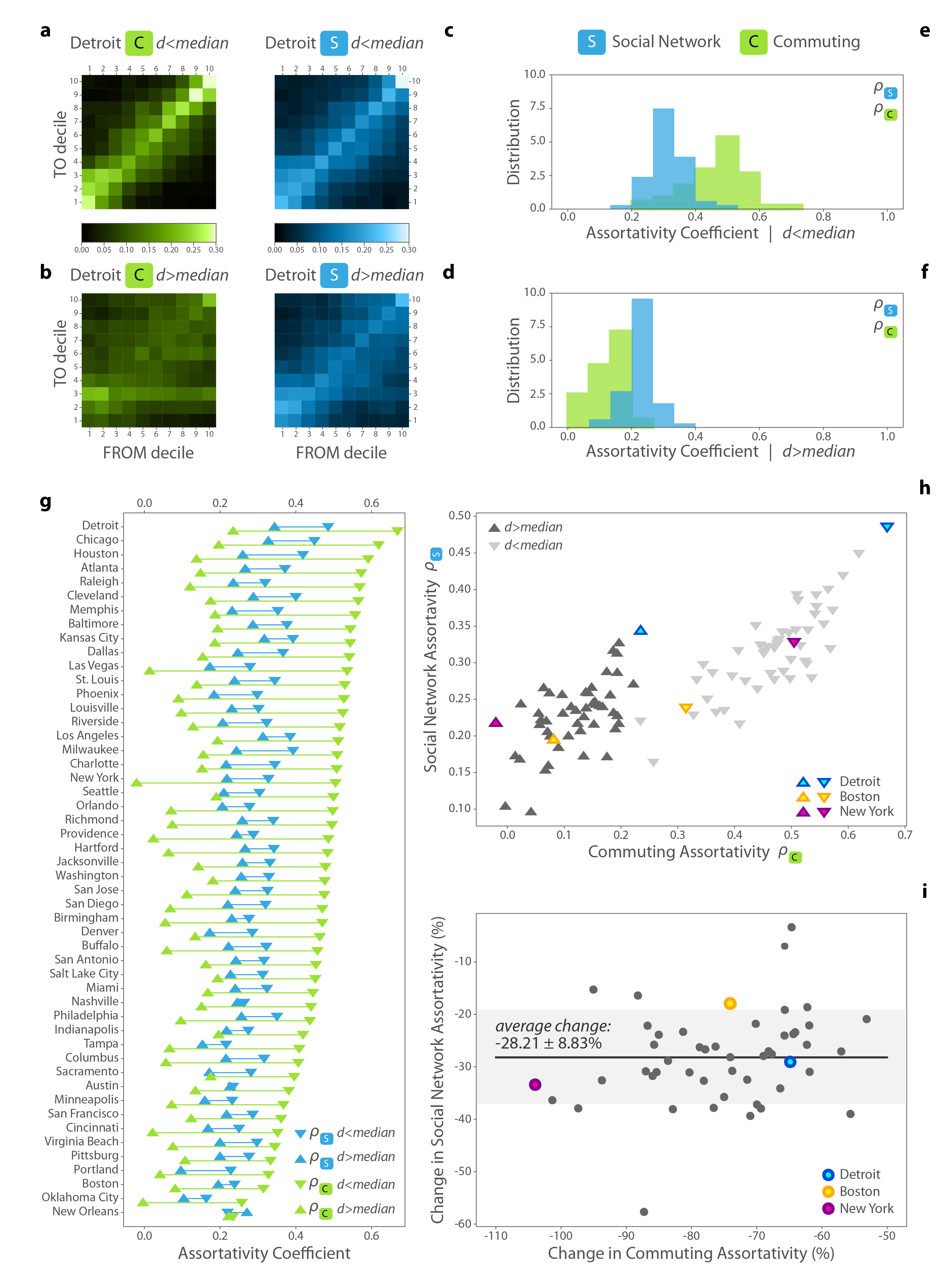}
\caption{\footnotesize Panels (a)-(d) show the $C$ and $S$ assortativity matrices for the below ($d<median$) and above median ($d>median$) commuting users in a selected metropolitan area, Detroit. (e)-(f) The corresponding distributions of $\rho_C$ (green) and $\rho_S$ (blue) for all 50 metropolitan areas for users with $d>{median}$ and $d<{median}$. (g) Pairwise values of $\rho_C$ and $\rho_S$ for users with $d>{median}$ and $d<{median}$ by metropolitan areas. Metropolitan areas are sorted in decreasing order by $\rho_C$ for easier representation. (h) Social network assortativity versus commuting assortativity for below and above median commuters with selected cities from Fig.~3 labeled. (i) Decrease in the commuting assortativity and the social network assortativity measured in percentage. Black horizontal line corresponds to the average change in social network assortativity. Grey shaded area marks the standard deviation.}
\label{fig:fig4}
\end{figure}

To investigate the association between long-distance commute and social mixing on the aggregate city-level in more detail, we separate the baseline sample of the $C$ and $S$ matrices by commuting distance. Thus, we create a $C$ and $S$ matrix from users commuting to a distance $d<{median}$ and $d>{median}$, as in the example in Figure~\ref{fig:fig4}a-d, where we show these four matrices (two for both $C$ and $S$) for Detroit. These matrices indicate that for users commuting an above median distance, matrices are less diagonal, and reflect more diverse and less segregated commuting and social connections. Panels (e) and (f) from Figure~\ref{fig:fig4} present the distributions of $\rho_C$ and $\rho_S$ for the two subgroups of users in all 50 metropolitan areas. As expected, longer commuting distance is associated with less assortativity because distant workplaces are likely to be located in socio-economically different environments as compared to home location. This might be due to spatial clustering of tracts with similar annual household incomes \parencite{Chodrow2017}, leading to shorter commute patterns landing in places with similar income level. In parallel, we observe that longer commutes are also associated with lower levels of assortativity of online social network ties such that off-diagonal social ties are relatively more likely for $d>{median}$ distances than for $d<{median}$. However, while $\rho_C$ falls sharply for $d>{median}$ distances compared to $d<{median}$, the difference of $\rho_S$ is moderate in Figure~\ref{fig:fig4}e-f. This finding indicates that although long-distance commutes can link disparate neighborhoods, not all of the diversity generated by commuting has imprints on social connections. Instead, income homophily remains a major yet weaker factor of social tie selection for long commuters as well.

Despite the heterogeneity of metro areas, results in Figure~\ref{fig:fig4}g show general patterns in two regards. First, assortativity of both commuting and social networks are lower for long-distance commuters in every metropolitan area. Second, the assortativity reduction between shorter and longer than median commutes is decreasing sharply, while the reduction of social network assortativity is moderate and takes similar values for every metropolitan area. The Pearson correlation coefficient between the two assortativity values $\rho_S$ and $\rho_C$ is 0.80 for short commuters and 0.72 for long commuters, thus they signify a strong relationship between mobility and social network assortativity patterns for both user groups (Figure~\ref{fig:fig4}h). 
To understand the magnitudes of change, we calculate the percentage of social network assortativity reduction by ($(\rho_{S,d>median}-\rho_{S,d<median}) / \rho_{S,d<median}$) and the percentage of commuting assortativity reduction by ($(\rho_{C,d>median}-\rho_{C,d<median}) / \rho_{C,d<median}$) for each city.  Illustrating these metrics, Figure~\ref{fig:fig4}i shows that the decrease in commuting assortativity ranges on a wide scale between -50\% and -100\%. However, the decrease in the social network assortativity concentrates around the average value of $-28\pm9\%$. Remarkably, this signals a universal pattern of social mixing potentials across very different urban settings and it explains a general trend of how mixing through commutes manifests in social inclusion. 
SI 7 illustrates that the uniform $\sim 30\%$ decrease disappears if we separate two user groups randomly instead of by commuting distance, but this observation remains consistent across multiple absolute distance thresholds (3~km, 5~km, and 10~km). In addition, in SI 8, we show that assortativity reduction by long-distance commute is a result of increasing social mixing of users from poorest and to some extent, from the richest neighborhoods.

\section{Discussion}

Understanding the complex behavioral patterns of people is crucial to develop more liveable, equal and sustainable urban environments. Our study contributes to this challenge by using large-scale geolocated Twitter data to study the 
role of commuting in the composition and assortativity of social interaction. 
We illustrate that long-distance commuting acts against structural closure and income homophily of social relationships and reduces segregation between remote income classes by facilitating connections and mixing. We show that home-work commutes and online social ties are not equally assortative in every metropolitan area, but in most cases, commuting is even more likely to point to places with similar income level than online social connections. Our findings suggest that longer commutes are more likely to connect places with different income levels, which contributes to the development of more diverse and less assortative social ties. Moreover, working further away from home results in more heterogeneous social connections in every metropolitan area. 

Our results suggest that urban mobility has a fundamental role in fulfilling the promise of social inclusion and reduction of social segregation in cities. The association between commuting distance and social networks is remarkably stable across all metropolitan areas with different size and spatial structure \parencite{Boeing2018}. This universal pattern highlights that commuting-enabled social mixing follows similar mechanisms regardless of the urban context. We find that facilitating the access between distant neighborhoods can reduce segregation in metropolitan areas, while gains in social inclusion are limited to a 30\% reduction of assortativity. These results signal that providing access across disparate neighborhoods cannot erase mechanisms of social network segregation but can mitigate the divide between rich and poor.


The methodology applied in this paper could easily be extended to other cities with large populations of geolocated Twitter users, and where granular census data with similar spatial resolution is available. However, this approach is not without limitations. While we are confident in our approach to identify home and work locations of users, we cannot confirm whether the identified work locations are actual workplaces or any another facility that people visit frequently during daytime (such as restaurants, schools, etc.). We measure commuting distances as the Euclidean distance between the home location and the work location, whereas in multiple cities, physical obstacles such as rivers might considerably increase to travel times or change the socio-economic segregation patterns of settlements \parencite{toth2019inequality}. We are not aware of the available modalities to reach work destinations, but we admit that it would also introduce a large variability into travel times. We choose this simplification because both travel times with a car or public transportation might depend on the exact time of the day and varying traffic conditions. Both the underestimation of commuting distances and the inclusion of users who might not have a regular workplace can result that the observed commuting in our case (see SI Figure 3) falls behind the commuting distances reported in the American Community Survey.
Because we do not use an absolute threshold to distinguish long and short commutes, and we use the city-wise median to divide the users into categories, we believe that the aforementioned biases do not affect our results drastically. However, we test both the results of Figure~2 and Figure~4i for different absolute distance thresholds, 3~km, 5~km and 10~km, where our results still hold (see SI 4 and SI 7).

Even though the fraction of users present in the analysis is proportional to the population size of the 50 metropolitan areas (see SI Figure~2), we have to highlight that our dataset is not representative for the US population and results have to be interpreted accordingly. \cite{Hargittai2011} finds that African American users are overrepresented on the platform, and Twitter users are predominantly young, well-educated \parencite{Webster2010, Sloan2015} and unrepresentative of other ethnicities \parencite{Mislove2011,Malik2015}. Therefore, we cannot generalize our findings to the whole population of these metropolitan areas. Another limitation of the study could be that the free 1\% sample from Twitter Streaming API was used for the initial data collection. \cite{Morstatter2013,Morstatter2014} confirms that tweets filtered to containing GPS coordinates are retrieved to almost 90\% of the time compared to the full dataset. By imposing strict count limits, spatio-temporal constraints and mutual followership for ties, we believe that our sample is less distorted from bot activity than what \cite{Pfeffer2018} would suggest.

Despite the imperfection of the data, we believe that the presented exercise offers useful insights to the structure of social connections within urban areas. Such large-scale, micro-level analysis enables us to uncover the fundamental patterns behind segregation, inequality or the lack of inclusion inside cities. Publicly available online social network data can complementing official census reports or surveys and can provide opportunities to detect and react to societal patterns and changes.


\section{Materials and methods}
\label{sec:mm}

\subsection{Data collection and combination methods}
We focus on users of the online social networking site of Twitter who posted tweets frequently containing precise geographic information. More specifically, we use a unique, historical database rich in tweets containing GPS coordinates \parencite{Dobos2013, Kondor2014}. These tweets originate from users who enabled the exact geolocation option on their smartphones. Overall, we detect the three most frequent tweeting locations of users as spatial clusters of their locations in the 50 most populated metropolitan areas of the United States. We use the Friend-of-Friend algorithm \parencite{huchra1982groups} to cluster the spatial coordinates for each user. This algorithm is a paralellizable, scalable clustering algorithm known from astronomy, and it is widely used to identify galaxy clusters \parencite{Kwon2010}. In our case, any two tweet coordinates of the same person are considered to belong to the same spatial cluster if their separation is less than 1~km. For each cluster, we determine the first two moments of the coordinate distribution. Before calculating the mean coordinates of the cluster, we trim data points until all points are inside a $3\sigma$ radius to eliminate outliers. We keep the aforementioned three highest cardinality clusters per user \parencite{Dobos2013, Kallus2015}. 

To determine the possible home and work locations of users, we follow the approach proposed by \cite{McNeill2017}. We assume that the home and work locations of users are within the previously detected three clusters. We select users for whom at least two out of the three clusters are within the same metropolitan area from the top 50 metropolitan areas of the United States and one of these clusters is their top cardinality location. First, we calculate the daily timeline of clusters for each user based on the timestamp of the tweets with hourly aggregation, converting all UTC tweet timestamps to local times across the whole US. We only consider users with more than 15 tweets on weekdays (Monday to Friday) in total. Local aggregated weekday timelines of two clusters for a sample user are presented in Figure~\ref{fig:fig1}a. We calculate the share of tweets sent between 9AM and 5PM on weekdays to capture messages predominantly sent during the working hours. Similarly, we calculate the share of tweets sent between 8PM and 8AM on weekdays contributing to a possible home tweeting fraction. Then, the cluster with the highest work tweet share or home tweet share becomes the work and home cluster of the user.

Commuting of users is characterized by the overhead distance between their home and work locations. We restricted our sample to users with at least 0.1~km commutes. Thus, we have 975,492 users in our sample. The distribution of observed commuting distances for each metro area are presented in SI 3. Additionally, we attach socio-economic data to each home and work location in the observed metropolitan areas from the 2012 American Community Survey. More precisely, we map the home locations of users into the census tracts of the top 50 US metropolitan areas and attribute the average annual household income of the census tract to each user living there. After that, we sort users into city-wise income deciles based on the average annual household incomes, and we apply the same approach to determine the average income and the income decile of their workplaces. Figure~\ref{fig:fig1}b shows the commute of the same sample user and the income level of the surrounding census tracts. 

Social connections of users are defined as their mutual followership relations on Twitter as they represent relative stronger ties in context of online social networks \parencite{Szule2014}. Figure~\ref{fig:fig1}c represents a sample ego network that we construct for every user from our home-work sample who has at least 1 mutual followership tie within the same metropolitan area. In the end, we have 348,850 users for whom we have both the home and work location information, and a mutual followership ego network. The composition and spatial distribution of our final sample is presented in SI 1. Through the home location of the user's friends, we can infer their income, thus, we are able to characterize the socio-economic status of the neighbors in the ego networks by identifying their income deciles. Figure~\ref{fig:fig1}c shows this characterization by using the same colorscale for both the ego and its first neighbors as the choropleth map in Figure~\ref{fig:fig1}b.

At the individual level, commuting and online social ties of our users are characterized by multiple different indicators. We measure user commutes by the Euclidean distance $d$ between their inferred home and work locations. We calculate degree and local clustering coefficient from their ego networks. We also measure the average income difference between their own home income and the home income of their friends, following the formula below: 

\begin{equation}
    \Delta I =\frac{1}{\textrm{\#neighbors}}\sum_{f\in\textrm{neighbors}} \log_{10}\left|I_f-I_{ego}\right|
    \label{eq:income}
\end{equation}

\subsection{Assortativity metrics}

At the aggregated, metropolitan area level, we create multiple different assortativity matrices between income deciles $D$  for each metropolitan area. First, an assortativity matrix of commuting is constructed, where we capture the probability $C_{ij}$ that a user $u$ belonging to a home census tract in income decile $D=i$ commutes to a tract with income decile $D=j$ to work. Second, we measure the conditional probabilities of social ties across home census tracts in different income deciles, the social network assortativity matrix $S$. The element $S_{ij}$ of this matrix measures the probability that a user $u$ from income decile $D=i$ has a mutual followership tie to a user in income decile $D=j$. Formally, the two matrices can be calculated as

\begin{align}\displaystyle
        C_{ij} &= \frac{%
            \sum\limits_{%
                \left\{%
                    u \in U\left| %
                        D_{u,\mathrm{home}=j},%
                        D_{u,\mathrm{work}=i}%
                    \right.%
                \right\}%
            }%
            1
        }{%
            \sum\limits_{%
                \left\{%
                    u \in U \left|%
                        D_{u,\mathrm{home}=j}%
                    \right.%
                \right\}%
            }%
            1%
        }\\
        S_{ij} &= \frac{%
            \sum\limits_{%
                \left\{
                    u \in U \left| D_{u,\mathrm{home}}=j\right.    
                \right\}
            }%
            \frac{1}{k_u}
            \sum\limits_{%
                \left\{%
                    e_{uf} \in E_u \left| %
                        D_{f,\mathrm{home}=i}%
                    \right.%
                \right\}%
            }%
            1
        }{%
            \sum\limits_{%
                \left\{%
                    u \in U\left|%
                        D_{u,\mathrm{home}=j}%
                    \right.%
                \right\}%
            }%
            1%
        },
    \label{eq:assort_matrices}
\end{align}
where $U$ is the user set within a metropolitan area for which we calculate the matrices, $E_u$ is the set of edges connected to the user $u$, $k_u$ is the degree of ego user $u$ in the ego network,  $e_uf$ is the undirected edge between user $u$ and $f$, $D_u$ and $D_f$ are the (home or work) deciles of users $u$ and $f$, respectively. We also measure two additional friendship and commuting assortativity matrices, $S^{d>\mathrm{median}}$, $S^{d<\mathrm{median}}$, $C^{d>\mathrm{median}}$ and $C^{d<\mathrm{median}}$, for users commuting more or less than the median commute in the given metropolitan area. In these cases, the set $U$ is what is different in the matrices from Eq.~\ref{eq:assort_matrices}.

We measure assortativity in these matrices by calculating the Pearson correlation coefficient $\rho$ of the matrix entries. If we normalize the elements of matrix $X$ such that $\tilde{X}_{ij} = X_{ij}/n$, where $n=\sum_{i,j}X_{ij}$, the sum of the elements of a matrix, then $\rho$ captures how diagonal these matrices are:

\newcommand{\sij}{\sum\limits_{i,j}}
\newcommand{\xij}{\tilde{X}_{ij}}

\begin{equation}{\displaystyle
    \rho_X = \frac{\sij ij \xij - \sij i \xij \sij j \xij}{\sqrt{\sij i^2 \xij - \left(\sij i \xij\right)^2}\sqrt{\sij j^2 \xij - \left(\sij j \xij\right)^2}}},
    \label{eq:rho}
\end{equation}
where the summation for $i$ and $j$ both go over all of the income deciles $D=1,\dots,10$. An assortativity value $\rho=+1$ would mean a completely diagonal, thus, completely assortative matrix, whereas $\rho\approx 0$ values indicate the lack of any preference for people following others from the very same income class of their own.

\newpage
 
\printbibliography

@article{Morales2019,
author = {Morales, Alfredo J. and Dong, Xiaowen and Bar-Yam, Yaneer and {‘Sandy' Pentland}, Alex},
doi = {10.1098/rsos.190573},
isbn = {0000000280},
issn = {2054-5703},
journal = {Royal Society Open Science},
month = {oct},
number = {10},
pages = {190573},
title = {{Segregation and polarization in urban areas}},
url = {https://royalsocietypublishing.org/doi/10.1098/rsos.190573},
volume = {6},
year = {2019}
}

@article{bokanyi2017prediction,
archivePrefix = {arXiv},
arxivId = {1703.07708},
author = {Bok{\'{a}}nyi, Eszter and L{\'{a}}bszki, Zolt{\'{a}}n and Vattay, G{\'{a}}bor},
doi = {10.1140/epjds/s13688-017-0112-x},
eprint = {1703.07708},
issn = {2193-1127},
journal = {EPJ Data Science},
month = {dec},
number = {1},
pages = {14},
title = {{Prediction of employment and unemployment rates from Twitter daily rhythms in the US}},
url = {http://arxiv.org/abs/1703.07708 http://epjdatascience.springeropen.com/articles/10.1140/epjds/s13688-017-0112-x},
volume = {6},
year = {2017}
}

@article{Norbutas2018,
author = {Norbutas, Lukas and Corten, Rense},
doi = {10.1016/j.socnet.2017.06.002},
issn = {03788733},
journal = {Social Networks},
month = {jan},
pages = {120--134},
publisher = {Elsevier B.V.},
title = {{Network structure and economic prosperity in municipalities: A large-scale test of social capital theory using social media data}},
url = {http://dx.doi.org/10.1016/j.socnet.2017.06.002 https://linkinghub.elsevier.com/retrieve/pii/S0378873316300375},
volume = {52},
year = {2018}
}

@article{Leo2016,
archivePrefix = {arXiv},
arxivId = {1612.04580},
author = {Leo, Yannick and Fleury, Eric and Alvarez-Hamelin, J. Ignacio and Sarraute, Carlos and Karsai, M{\'{a}}rton},
doi = {10.1098/rsif.2016.0598},
eprint = {1612.04580},
issn = {1742-5689},
journal = {Journal of The Royal Society Interface},
month = {dec},
number = {125},
pages = {20160598},
pmid = {27974571},
title = {{Socioeconomic correlations and stratification in social-communication networks}},
url = {https://royalsocietypublishing.org/doi/10.1098/rsif.2016.0598},
volume = {13},
year = {2016}
}

@article{Sloan2015,
author = {Sloan, Luke and Morgan, Jeffrey and Burnap, Pete and Williams, Matthew},
doi = {10.1371/journal.pone.0115545},
editor = {Preis, Tobias},
issn = {1932-6203},
journal = {PLOS ONE},
month = {mar},
number = {3},
pages = {e0115545},
title = {{Who Tweets? Deriving the Demographic Characteristics of Age, Occupation and Social Class from Twitter User Meta-Data}},
url = {http://dx.plos.org/10.1371/journal.pone.0115545 https://dx.plos.org/10.1371/journal.pone.0115545},
volume = {10},
year = {2015}
}

@inproceedings{Kallus2017,
address = {Cham},
author = {Kallus, Zs{\'{o}}fia and Kondor, D{\'{a}}niel and St{\'{e}}ger, J{\'{o}}zsef and Csabai, Istv{\'{a}}n and Bok{\'{a}}nyi, Eszter and Vattay, G{\'{a}}bor},
booktitle = {ICT Innovations 2017: Data-Driven Innovation},
doi = {10.1007/978-3-319-67597-8_1},
editor = {Trajanov, Dimitar and Bakeva, Verica},
isbn = {978-3-319-67597-8},
issn = {18650929},
pages = {3--12},
publisher = {Springer International Publishing},
title = {{Video Pandemics: Worldwide Viral Spreading of Psy's Gangnam Style Video}},
url = {https://doi.org/10.1007/978-3-319-67597-8{\_}1 http://link.springer.com/10.1007/978-3-319-67597-8{\_}1},
volume = {778},
year = {2017}
}

@article{Bokanyi2016,
archivePrefix = {arXiv},
arxivId = {1605.02951},
author = {Bok{\'{a}}nyi, Eszter and Kondor, D{\'{a}}niel and Dobos, L{\'{a}}szl{\'{o}} and Sebők, Tam{\'{a}}s and St{\'{e}}ger, J{\'{o}}zsef and Csabai, Istv{\'{a}}n and Vattay, G{\'{a}}bor},
doi = {10.1057/palcomms.2016.10},
eprint = {1605.02951},
issn = {2055-1045},
journal = {Palgrave Communications},
month = {dec},
number = {1},
pages = {16010},
publisher = {Nature Publishing Group},
title = {{Race, religion and the city: twitter word frequency patterns reveal dominant demographic dimensions in the United States}},
url = {http://www.palgrave-journals.com/articles/palcomms201610 http://www.nature.com/articles/palcomms201610},
volume = {2},
year = {2016}
}

@inproceedings{Dobos2013,
archivePrefix = {arXiv},
arxivId = {1311.0841},
author = {Dobos, Laszlo and Szule, Janos and Bodnar, Tamas and Hanyecz, Tamas and Sebok, Tamas and Kondor, Daniel and Kallus, Zsofia and Steger, Jozsef and Csabai, Istvan and Vattay, Gabor},
booktitle = {2013 IEEE 4th International Conference on Cognitive Infocommunications (CogInfoCom)},
doi = {10.1109/CogInfoCom.2013.6719259},
eprint = {1311.0841},
isbn = {978-1-4799-1546-0},
month = {dec},
pages = {289--294},
publisher = {IEEE},
title = {{A multi-terabyte relational database for geo-tagged social network data}},
url = {http://ieeexplore.ieee.org/document/6719259/},
year = {2013}
}

@article{Lambiotte2008,
archivePrefix = {arXiv},
arxivId = {0802.2178},
author = {Lambiotte, Renaud and Blondel, Vincent D. and de Kerchove, Cristobald and Huens, Etienne and Prieur, Christophe and Smoreda, Zbigniew and {Van Dooren}, Paul},
doi = {10.1016/j.physa.2008.05.014},
eprint = {0802.2178},
isbn = {0378-4371},
issn = {03784371},
journal = {Physica A: Statistical Mechanics and its Applications},
month = {sep},
number = {21},
pages = {5317--5325},
title = {{Geographical dispersal of mobile communication networks}},
url = {https://linkinghub.elsevier.com/retrieve/pii/S0378437108004342},
volume = {387},
year = {2008}
}

@article{Webster2010,
author = {Webster, Tom},
journal = {Edison Research/ Arbitron Internet and Multimedia Study.},
title = {{Twitter Usage In America : 2010}},
url = {http://www.onecommunity.org/wp-content/uploads/2010/04/Twitter{\_}Usage{\_}In{\_}America{\_}2010.pdf},
year = {2010}
}

@article{Sampson2008,
archivePrefix = {arXiv},
arxivId = {NIHMS150003},
author = {Sampson, Robert J.},
doi = {10.1086/589843},
eprint = {NIHMS150003},
isbn = {6176321972},
issn = {0002-9602},
journal = {American Journal of Sociology},
month = {jul},
number = {1},
pages = {189--231},
pmid = {1000000221},
title = {{Moving to Inequality: Neighborhood Effects and Experiments Meet Social Structure}},
url = {http://www.journals.uchicago.edu/doi/10.1086/589843},
volume = {114},
year = {2008}
}

@article{McNeill2017,
author = {McNeill, Graham and Bright, Jonathan and Hale, Scott A.},
doi = {10.1140/epjds/s13688-017-0120-x},
issn = {2193-1127},
journal = {EPJ Data Science},
month = {dec},
number = {1},
pages = {24},
publisher = {SpringerOpen},
title = {{Estimating local commuting patterns from geolocated Twitter data}},
url = {http://epjdatascience.springeropen.com/articles/10.1140/epjds/s13688-017-0120-x},
volume = {6},
year = {2017}
}

@article{Calabrese2011,
archivePrefix = {arXiv},
arxivId = {1101.4505},
author = {Calabrese, Francesco and Smoreda, Zbigniew and Blondel, Vincent D. and Ratti, Carlo},
doi = {10.1371/journal.pone.0020814},
editor = {Scalas, Enrico},
eprint = {1101.4505},
issn = {1932-6203},
journal = {PLoS ONE},
month = {jul},
number = {7},
pages = {e20814},
title = {{Interplay between Telecommunications and Face-to-Face Interactions: A Study Using Mobile Phone Data}},
url = {https://dx.plos.org/10.1371/journal.pone.0020814},
volume = {6},
year = {2011}
}

@inproceedings{Mislove2011,
author = {Mislove, Alan and Lehmann, Sune and Ahn, Yong-yeol and Onnela, Jukka-Pekka and Rosenquist, J Niels},
booktitle = {Int'l AAAI Conference on Weblogs and Social Media (ICWSM)},
pages = {554--557},
title = {{Understanding the Demographics of Twitter Users}},
url = {http://www.aaai.org/ocs/index.php/ICWSM/ICWSM11/paper/viewFile/2816/3234},
year = {2011}
}

@article{Nieuwenhuis2020,
author = {Nieuwenhuis, Jaap and Tammaru, Tiit and van Ham, Maarten and Hedman, Lina and Manley, David},
doi = {10.1177/0042098018807628},
issn = {0042-0980},
journal = {Urban Studies},
month = {jan},
number = {1},
pages = {176--197},
publisher = {SAGE Publications Ltd},
title = {{Does segregation reduce socio-spatial mobility? Evidence from four European countries with different inequality and segregation contexts}},
url = {http://journals.sagepub.com/doi/10.1177/0042098018807628},
volume = {57},
year = {2020}
}

@article{Viry2012,
author = {Viry, Gil},
doi = {10.1016/j.socnet.2011.07.003},
issn = {03788733},
journal = {Social Networks},
month = {jan},
number = {1},
pages = {59--72},
publisher = {Elsevier B.V.},
title = {{Residential mobility and the spatial dispersion of personal networks: Effects on social support}},
url = {http://dx.doi.org/10.1016/j.socnet.2011.07.003 https://linkinghub.elsevier.com/retrieve/pii/S0378873311000475},
volume = {34},
year = {2012}
}

@incollection{Morstatter2013,
archivePrefix = {arXiv},
arxivId = {1306.5204},
author = {Joseph, Kenneth and Landwehr, Peter M. and Carley, Kathleen M.},
booktitle = {Association fo the Advanced of Artificial Intelligence},
doi = {10.1007/978-3-319-05579-4_10},
eprint = {1306.5204},
isbn = {9783319055787},
issn = {16113349},
month = {jun},
pages = {75--83},
pmid = {27340949},
title = {{Two 1{\%}s Don't Make a Whole: Comparing Simultaneous Samples from Twitter's Streaming API}},
url = {http://arxiv.org/abs/1306.5204 http://link.springer.com/10.1007/978-3-319-05579-4{\_}10},
year = {2014}
}

@article{Heine2021,
author = {Heine, Cate and Marquez, Cristina and Santi, Paolo and Sundberg, Marcus and Nordfors, Miriam and Ratti, Carlo},
doi = {10.1371/journal.pone.0247996},
isbn = {1111111111},
issn = {19326203},
journal = {PLoS ONE},
number= {3 March},
pages = {1--14},
title = {{Analysis of mobility homophily in Stockholm based on social network data}},
url = {http://dx.doi.org/10.1371/journal.pone.0247996},
volume = {16},
year = {2021}
}

@article{Watts1998,
author = {Watts, Duncan J. and Strogatz, Steven H.},
doi = {10.1038/30918},
issn = {00280836},
journal = {Nature},
month = {jun},
number = {6684},
pages = {440--442},
pmid = {9623998},
publisher = {Macmillan Magazines Ltd},
title = {{Collective dynamics of 'small-world9 networks}},
url = {http://us.imdb.com},
volume = {393},
year = {1998}
}

@inproceedings{Kondor2014,
address = {New York, New York, USA},
author = {Kondor, D{\'{a}}niel and Dobos, L{\'{a}}szl{\'{o}} and Csabai, Istv{\'{a}}n and Bodor, Andr{\'{a}}s and Vattay, G{\'{a}}bor and Budav{\'{a}}ri, Tam{\'{a}}s and Szalay, Alexander S},
booktitle = {Proceedings of the 26th International Conference on Scientific and Statistical Database Management - SSDBM '14},
doi = {10.1145/2618243.2618245},
isbn = {9781450327220},
pages = {1--4},
publisher = {ACM Press},
title = {{Efficient classification of billions of points into complex geographic regions using hierarchical triangular mesh}},
url = {http://dl.acm.org/citation.cfm?doid=2618243.2618245},
year = {2014}
}

@article{bettencourt2013origins,
author = {Bettencourt, Lu{\'{i}}s M A},
doi = {10.1126/science.1235823},
isbn = {1095-9203 (Electronic)$\backslash$r0036-8075 (Linking)},
issn = {0036-8075},
journal = {Science},
month = {jun},
number = {6139},
pages = {1438--1441},
pmid = {23788793},
title = {{The Origins of Scaling in Cities}},
url = {http://www.ncbi.nlm.nih.gov/pubmed/23788793 http://www.sciencemag.org/cgi/doi/10.1126/science.1235823 https://www.sciencemag.org/lookup/doi/10.1126/science.1235823},
volume = {340},
year = {2013}
}

@article{Szule2014,
author = {Sz{\"{u}}le, J{\'{a}}nos and Kondor, D{\'{a}}niel and Dobos, L{\'{a}}szl{\'{o}} and Csabai, Istv{\'{a}}n and Vattay, G{\'{a}}bor},
doi = {10.1371/journal.pone.0111973},
editor = {Garcia-Ojalvo, Jordi},
issn = {1932-6203},
journal = {PLoS ONE},
month = {nov},
number = {11},
pages = {e111973},
pmid = {25383796},
title = {{Lost in the City: Revisiting Milgram's Experiment in the Age of Social Networks}},
url = {http://www.ncbi.nlm.nih.gov/pubmed/25383796 https://dx.plos.org/10.1371/journal.pone.0111973},
volume = {9},
year = {2014}
}

@article{Kallus2015,
author = {Kallus, Zs{\'{o}}fia and Barankai, Norbert and Sz{\"{u}}le, J{\'{a}}nos and Vattay, G{\'{a}}bor},
doi = {10.1371/journal.pone.0126713},
editor = {Jiang, Bin},
issn = {1932-6203},
journal = {PLOS ONE},
month = {may},
number = {5},
pages = {e0126713},
title = {{Spatial Fingerprints of Community Structure in Human Interaction Network for an Extensive Set of Large-Scale Regions}},
url = {http://dx.plos.org/10.1371/journal.pone.0126713 https://dx.plos.org/10.1371/journal.pone.0126713},
volume = {10},
year = {2015}
}

@unpublished{Blumenstock2019,
author = {Blumenstock, Joshua and Chi, Guanghua and Tan, Xu},
booktitle = {CEPR Discussion Papers},
title = {{Migration and the Value of Social Networks}},
year = {2019}
}

@article{Hargittai2011,
author = {Hargittai, Eszter and Litt, Eden},
doi = {10.1177/1461444811405805},
isbn = {1461-4448},
issn = {1461-4448},
journal = {New Media {\&} Society},
month = {aug},
number = {5},
pages = {824--842},
title = {{The tweet smell of celebrity success: Explaining variation in Twitter adoption among a diverse group of young adults}},
url = {http://nms.sagepub.com/content/13/5/824.full.pdf http://journals.sagepub.com/doi/10.1177/1461444811405805},
volume = {13},
year = {2011}
}

@article{Bailey2020,
address = {Cambridge, MA},
author = {Bailey, Michael and Farrell, Patrick and Kuchler, Theresa and Stroebel, Johannes},
doi = {10.1016/j.jue.2020.103264},
institution = {National Bureau of Economic Research},
issn = {00941190},
journal = {Journal of Urban Economics},
month = {jul},
pages = {103264},
title = {{Social connectedness in urban areas}},
url = {http://www.nber.org/papers/w26029{\%}0Ahttp://www.nber.org/papers/w26029.pdf http://www.nber.org/papers/w26029.pdf https://linkinghub.elsevier.com/retrieve/pii/S0094119020300358},
volume = {118},
year = {2020}
}

@article{FloridaMellander2015,
author = {Florida, Richard and Mellander, Charlotta},
journal = {Martin Prosperity Institute},
title = {{Segregated city: The geography of economic segregation in America's metros}},
url = {https://www.diva-portal.org/smash/get/diva2:868382/FULLTEXT01.pdf},
year = {2015}
}

@article{FryTaylor2012,
author = {Fry, Richard and Taylor, Paul},
journal = {Pew Research Center},
title = {{The rise of residential segregation by income}},
url = {https://www.pewresearch.org/wp-content/uploads/sites/3/2012/08/Rise-of-Residential-Income-Segregation-2012.2.pdf},
year = {2012}
}

@article{Massey1988,
author = {Massey, Douglas S and Denton, Nancy A},
journal = {Social Forces},
number = {2},
pages = {281--315},
title = {{The dimension of residential segregation}},
volume = {67},
year = {1988}
}

@article{Yip2016,
author = {Yip, Ngai Ming and Forrest, Ray and Xian, Shi},
doi = {10.1016/j.cities.2016.02.003},
issn = {02642751},
journal = {Cities},
pages = {156--163},
title = {{Exploring segregation and mobilities: Application of an activity tracking app on mobile phone}},
volume = {59},
year = {2016}
}

@article{Dahlin2008,
author = {Dahlin, Eric and Kelly, Erin and Moen, Phyllis},
doi = {10.1111/j.1533-8525.2008.00133.x},
issn = {00380253},
journal = {Sociological Quarterly},
number = {4},
pages = {719--736},
title = {{Is work the new neighborhood? Social ties in the workplace, family, and neighborhood}},
volume = {49},
year = {2008}
}

@article{Chodrow2017,
author = {Chodrow, Philip S},
doi = {10.1073/pnas.1708201114},
issn = {10916490},
journal = {Proceedings of the National Academy of Sciences of the United States of America},
number = {44},
pages = {11591--11596},
pmid = {29078323},
title = {{Structure and information in spatial segregation}},
volume = {114},
year = {2017}
}

@article{toth2019inequality,
author = {T{\'{o}}th, Gergő and Wachs, Johannes and {Di Clemente}, Riccardo and Jakobi, {\'{A}}kos and S{\'{a}}gv{\'{a}}ri, Bence and Kert{\'{e}}sz, J{\'{a}}nos and Lengyel, Bal{\'{a}}zs},
doi = {10.1038/s41467-021-21465-0},
issn = {2041-1723},
journal = {Nature Communications},
month = {dec},
number = {1},
pages = {1143},
title = {{Inequality is rising where social network segregation interacts with urban topology}},
url = {http://www.nature.com/articles/s41467-021-21465-0},
volume = {12},
year = {2021}
}

@article{Kwon2010,
author = {Kwon, Yongchul and Nunley, Dylan and Gardner, Jeffrey P and Balazinska, Magdalena and Howe, Bill and Loebman, Sarah},
doi = {10.1007/978-3-642-13818-8_11},
isbn = {3642138179},
issn = {03029743},
journal = {Lecture Notes in Computer Science (including subseries Lecture Notes in Artificial Intelligence and Lecture Notes in Bioinformatics)},
pages = {132--150},
title = {{Scalable clustering algorithm for N-body simulations in a shared-nothing cluster}},
url = {http://link.springer.com/10.1007/978-3-642-13818-8{\_}11},
volume = {6187 LNCS},
year = {2010}
}

@article{huchra1982groups,
author = {Huchra, J P and Geller, M. J.},
doi = {10.1086/160000},
issn = {0004-637X},
journal = {The Astrophysical Journal},
month = {jun},
pages = {423},
title = {{Groups of galaxies. I - Nearby groups}},
url = {http://adsabs.harvard.edu/doi/10.1086/160000},
volume = {257},
year = {1982}
}

@article{Wang2018,
author = {Wang, Qi and Phillips, Nolan Edward and Small, Mario L and Sampson, Robert J},
doi = {10.1073/pnas.1802537115},
issn = {0027-8424},
journal = {Proceedings of the National Academy of Sciences},
month = {jul},
number = {30},
pages = {7735--7740},
title = {{Urban mobility and neighborhood isolation in America's 50 largest cities}},
url = {http://www.pnas.org/lookup/doi/10.1073/pnas.1802537115},
volume = {115},
year = {2018}
}

@article{Dong2019,
archivePrefix = {arXiv},
arxivId = {1911.04027},
author = {Dong, Xiaowen and Morales, Alfredo J and Jahani, Eaman and Moro, Esteban and Lepri, Bruno and Bozkaya, Burcin and Sarraute, Carlos and Bar-Yam, Yaneer and Pentland, Alex},
doi = {10.1140/epjds/s13688-020-00238-7},
eprint = {1911.04027},
issn = {2193-1127},
journal = {EPJ Data Science},
month = {dec},
number = {1},
pages = {20},
title = {{Segregated interactions in urban and online space}},
url = {http://arxiv.org/abs/1911.04027 https://epjdatascience.springeropen.com/articles/10.1140/epjds/s13688-020-00238-7},
volume = {9},
year = {2020}
}

@article{Boeing2018,
author = {Boeing, Geoff},
doi = {10.1007/s41109-019-0189-1},
isbn = {4110901901},
issn = {2364-8228},
journal = {Applied Network Science},
month = {dec},
number = {1},
pages = {67},
publisher = {Applied Network Science},
title = {{Urban spatial order: street network orientation, configuration, and entropy}},
url = {https://appliednetsci.springeropen.com/articles/10.1007/s41109-019-0189-1},
volume = {4},
year = {2019}
}

@incollection{Ham2018,
author = {van Ham, Maarten and Tammaru, Tiit and Janssen, Heleen J},
booktitle = {Divided Cities},
doi = {10.1787/9789264300385-8-en},
number = {8774},
pages = {135--153},
publisher = {OECD},
title = {{A multi-level model of vicious circles of socio-economic segregation}},
url = {https://www.oecd-ilibrary.org/urban-rural-and-regional-development/divided-cities/a-multi-level-model-of-vicious-circles-of-socio-economic-segregation{\_}9789264300385-8-en},
volume = {615159},
year = {2018}
}

@article{Bora2014,
author = {Bora, Nibir and Chang, Yu-Han and Maheswaran, Rajiv},
doi = {10.1007/978-3-319-05579-4_2},
journal = {Proceedings of the international conference on social computing, behavioral-cultural modeling, and prediction},
pages = {11--18},
title = {{Mobility Patterns and User Dynamics in Racially Segregated Geographies of US Cities}},
url = {http://link.springer.com/10.1007/978-3-319-05579-4{\_}2},
year = {2014}
}

@article{Malik2015,
author = {Malik, Momin M and Lamba, Hemank and Nakos, Constantine and Pfeffer, J{\"{u}}rgen},
isbn = {9781577357360},
journal = {AAAI Workshop - Technical Report},
pages = {18--27},
title = {{Population bias in geotagged tweets}},
volume = {WS-15-18},
year = {2015}
}

@article{Pfeffer2018,
author = {Pfeffer, J{\"{u}}rgen and Mayer, Katja and Morstatter, Fred},
doi = {10.1140/epjds/s13688-018-0178-0},
issn = {2193-1127},
journal = {EPJ Data Science},
month = {dec},
number = {1},
pages = {50},
publisher = {The Author(s)},
title = {{Tampering with Twitter's Sample API}},
url = {http://dx.doi.org/10.1140/epjds/s13688-018-0178-0 https://epjdatascience.springeropen.com/articles/10.1140/epjds/s13688-018-0178-0},
volume = {7},
year = {2018}
}

@inproceedings{Morstatter2014,
address = {New York, New York, USA},
archivePrefix = {arXiv},
arxivId = {1401.7909},
author = {Morstatter, Fred and Pfeffer, J{\"{u}}rgen and Liu, Huan},
booktitle = {Proceedings of the 23rd International Conference on World Wide Web - WWW '14 Companion},
doi = {10.1145/2567948.2576952},
eprint = {1401.7909},
isbn = {9781450327459},
month = {jan},
pages = {555--556},
publisher = {ACM Press},
title = {{When is it biased?}},
url = {http://arxiv.org/abs/1401.7909 http://dl.acm.org/citation.cfm?doid=2567948.2576952},
year = {2014}
}

@book{jacobs2016death,
author = {Jacobs, Jane},
publisher = {Vintage},
title = {{The death and life of great American cities}},
year = {2016}
}

@article{glaeser2011cities,
author = {Glaeser, Edward},
journal = {Science},
number = {6042},
pages = {592--594},
publisher = {American Association for the Advancement of Science},
title = {{Cities, productivity, and quality of life}},
volume = {333},
year = {2011}
}

@article{duranton2020economics,
author = {Duranton, Gilles and Puga, Diego},
doi = {10.1257/jep.34.3.3},
issn = {19447965},
journal = {Journal of Economic Perspectives},
number = {3},
pages = {3--26},
title = {{The economics of urban density}},
volume = {34},
year = {2020}
}

@article{storper2004buzz,
author = {Storper, Michael and Venables, Anthony J},
doi = {10.1093/jnlecg/lbh027},
issn = {14682702},
journal = {Journal of Economic Geography},
number = {4},
pages = {351--370},
publisher = {Oxford University Press},
title = {{Buzz: Face-to-face contact and the urban economy}},
volume = {4},
year = {2004}
}

@article{chong2020economic,
author = {Chong, Shi Kai and Bahrami, Mohsen and Chen, Hao and Balcisoy, Selim and Bozkaya, Burcin and Pentland, Alex ‘Sandy'},
doi = {10.1140/epjds/s13688-020-00234-x},
issn = {21931127},
journal = {EPJ Data Science},
number = {1},
pages = {17},
publisher = {Springer Berlin Heidelberg},
title = {{Economic outcomes predicted by diversity in cities}},
volume = {9},
year = {2020}
}

@article{calabrese2011interplay,
archivePrefix = {arXiv},
arxivId = {1101.4505},
author = {Calabrese, Francesco and Smoreda, Zbigniew and Blondel, Vincent D and Ratti, Carlo},
doi = {10.1371/journal.pone.0020814},
eprint = {1101.4505},
issn = {19326203},
journal = {PLoS ONE},
number = {7},
pages = {e20814},
pmid = {21765888},
publisher = {Public Library of Science},
title = {{Interplay between telecommunications and face-to-face interactions: A study using mobile phone data}},
volume = {6},
year = {2011}
}

@article{glaeser2009inequality,
author = {Glaeser, Edward L and Resseger, Matt and Tobio, Kristina},
doi = {10.1111/j.1467-9787.2009.00627.x},
issn = {00224146},
journal = {Journal of Regional Science},
number = {4},
pages = {617--646},
publisher = {Wiley Online Library},
title = {{Inequality in cities}},
volume = {49},
year = {2009}
}

@article{ananat2011wrong,
author = {Ananat, Elizabeth Oltmans},
doi = {10.1257/app.3.2.34},
issn = {19457782},
journal = {American Economic Journal: Applied Economics},
number = {2},
pages = {34--66},
title = {{The wrong side(s) of the tracks: The causal effects of racial segregation on urban poverty and inequality}},
volume = {3},
year = {2011}
}

@article{toth2021inequality,
archivePrefix = {arXiv},
arxivId = {1909.11414},
author = {T{\'{o}}th, Gergő and Wachs, Johannes and Clemente, Riccardo Di and Jakobi, {\'{A}}kos and S{\'{a}}gv{\'{a}}ri, Bence and Kert{\'{e}}sz, J{\'{a}}nos and Lengyel, Bal{\'{a}}zs},
doi = {10.1038/s41467-021-21465-0},
eprint = {1909.11414},
issn = {23318422},
journal = {arXiv},
number = {1},
pages = {1--9},
publisher = {Nature Publishing Group},
title = {{Inequality is rising where social network segregation interacts with urban topology}},
volume = {12},
year = {2019}
}

@article{eagle2010network,
author = {Eagle, Nathan and Macy, Michael and Claxton, Rob},
doi = {10.1126/science.1186605},
issn = {00368075},
journal = {Science},
number = {5981},
pages = {1029--1031},
pmid = {20489022},
publisher = {American Association for the Advancement of Science},
title = {{Network diversity and economic development}},
volume = {328},
year = {2010}
}

@inproceedings{pappalardo2015using,
author = {Pappalardo, Luca and Pedreschi, Dino and Smoreda, Zbigniew and Giannotti, Fosca},
booktitle = {Proceedings - 2015 IEEE International Conference on Big Data, IEEE Big Data 2015},
doi = {10.1109/BigData.2015.7363835},
isbn = {9781479999255},
organization = {IEEE},
pages = {871--878},
title = {{Using big data to study the link between human mobility and socio-economic development}},
year = {2015}
}

@article{jiang2016timegeo,
author = {Jiang, Shan and Yang, Yingxiang and Gupta, Siddharth and Veneziano, Daniele and Athavale, Shounak and Gonz{\'{a}}lez, Marta C},
doi = {10.1073/pnas.1524261113},
issn = {10916490},
journal = {Proceedings of the National Academy of Sciences of the United States of America},
number = {37},
pages = {E5370--E5378},
pmid = {27573826},
publisher = {National Acad Sciences},
title = {{The TimeGeo modeling framework for urban motility without travel surveys}},
volume = {113},
year = {2016}
}

@inproceedings{florez2018measuring,
author = {Florez, Manuel A and Jiang, Shan and Li, Ruiqi and Rios, Ramiro A and Gonz{\'{a}}lez, Marta C},
booktitle = {Transportation Research Board, 96th Annual Meeting},
number = {03745},
title = {{Measuring the impacts of economic well being in commuting networks — A case study of Columbia}},
volume = {17},
year = {2016}
}

@article{roberto2015spatial,
archivePrefix = {arXiv},
arxivId = {1509.03678},
author = {Roberto, Elizabeth},
doi = {10.1177/0081175018796871},
eprint = {1509.03678},
issn = {14679531},
journal = {Sociological Methodology},
number = {1},
pages = {182--224},
title = {{The Spatial Proximity and Connectivity Method for Measuring and Analyzing Residential Segregation}},
volume = {48},
year = {2018}
}

@article{Dannemann2018,
author = {Dannemann, Teodoro and Sotomayor-G{\'{o}}mez, Boris and Samaniego, Horacio},
doi = {10.1098/rsos.180749},
issn = {2054-5703},
journal = {Royal Society Open Science},
month = {oct},
number = {10},
pages = {180749},
title = {{The time geography of segregation during working hours}},
url = {https://royalsocietypublishing.org/doi/10.1098/rsos.180749},
volume = {5},
year = {2018}
}

@article{Small2019,
author = {Small, Mario L. and Adler, Laura},
doi = {10.1146/annurev-soc-073018-022707},
issn = {03600572},
journal = {Annual Review of Sociology},
pages = {111--132},
title = {{The Role of Space in the Formation of Social Ties}},
volume = {45},
year = {2019}
}

@article{Eagle2009,
author = {Eagle, Nathan and Pentland, Alex and Lazer, David},
doi = {10.1073/pnas.0900282106},
issn = {00278424},
journal = {Proceedings of the National Academy of Sciences of the United States of America},
number = {36},
pages = {15274--15278},
pmid = {19706491},
title = {{Inferring friendship network structure by using mobile phone data}},
volume = {106},
year = {2009}
}

@article{abitbol2020interpretable,
  title={Interpretable socioeconomic status inference from aerial imagery through urban patterns},
  author={Abitbol, Jacob Levy and Karsai, Marton},
  journal={Nature Machine Intelligence},
  volume={2},
  number={11},
  pages={684--692},
  year={2020},
  publisher={Nature Publishing Group}
}

\newpage

\section{Acknowledgements}

Eszter Bok\'anyi was supported by the ÚNKP-20-4 New National Excellence Program of the Ministry for Innovation and Technology from the source of the National Research, Development and Innovation Fund of Hungary. M\'arton Karsai acknowledges support from the H2020 SoBigData++ project (H2020-871042) and the DataRedux ANR project (ANR-19-CE46-0008). Balázs Lengyel and Sándor Juhász acknowledge support from the Hungarian Scientific Research Fund (OTKA K-138970). We thank for the usage of ELKH Cloud (\url{https://science-cloud.hu/}) that significantly helped us achieving the results published in this paper. We thank József Stéger for helping in the maintenance of the Twitter database, and Szabolcs Tóth-Zs. (\url{https://bandart.eu/}) for figure design.

\cleardoublepage

\section*{Supplementary information}
\subsection*{SI 1: Observed users across the top 50 US metropolitan areas}

\begin{figure}[htb]
\centering
\includegraphics[width=145mm]{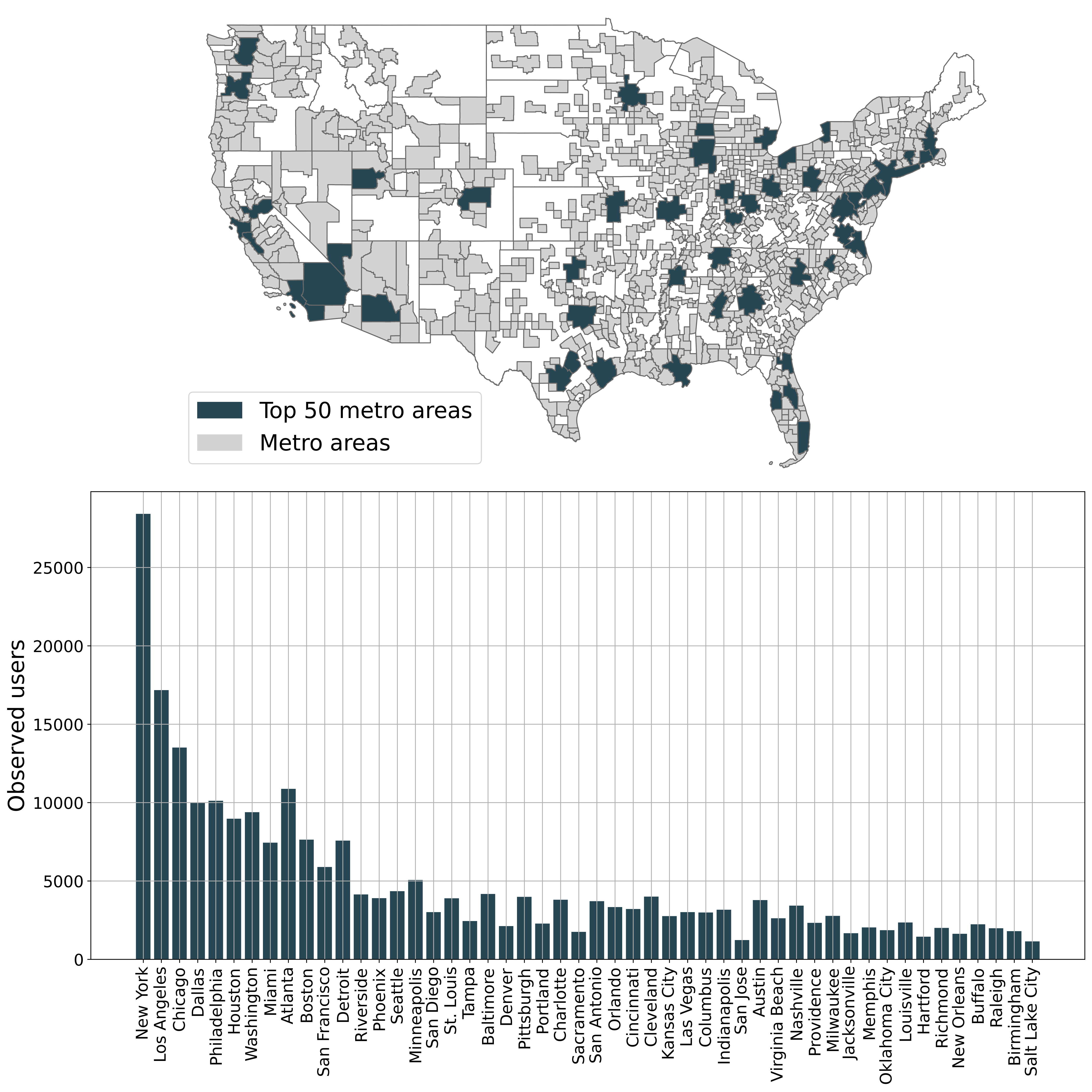}
\caption{\footnotesize 
(A) Map of the selected 50 metropolitan areas with the highest population in the US. (B) The histogram represents the number of observed users with home and work locations, minimum 100 meter commute and minimum 1 connection to a user with discovered home and work locations in the same metro area. The metro areas are ordered by population.
}
\end{figure}

\newpage

\subsection*{SI 2: Population and observed users in metro areas}

\begin{figure}[htb]
\centering
\includegraphics[width=70mm]{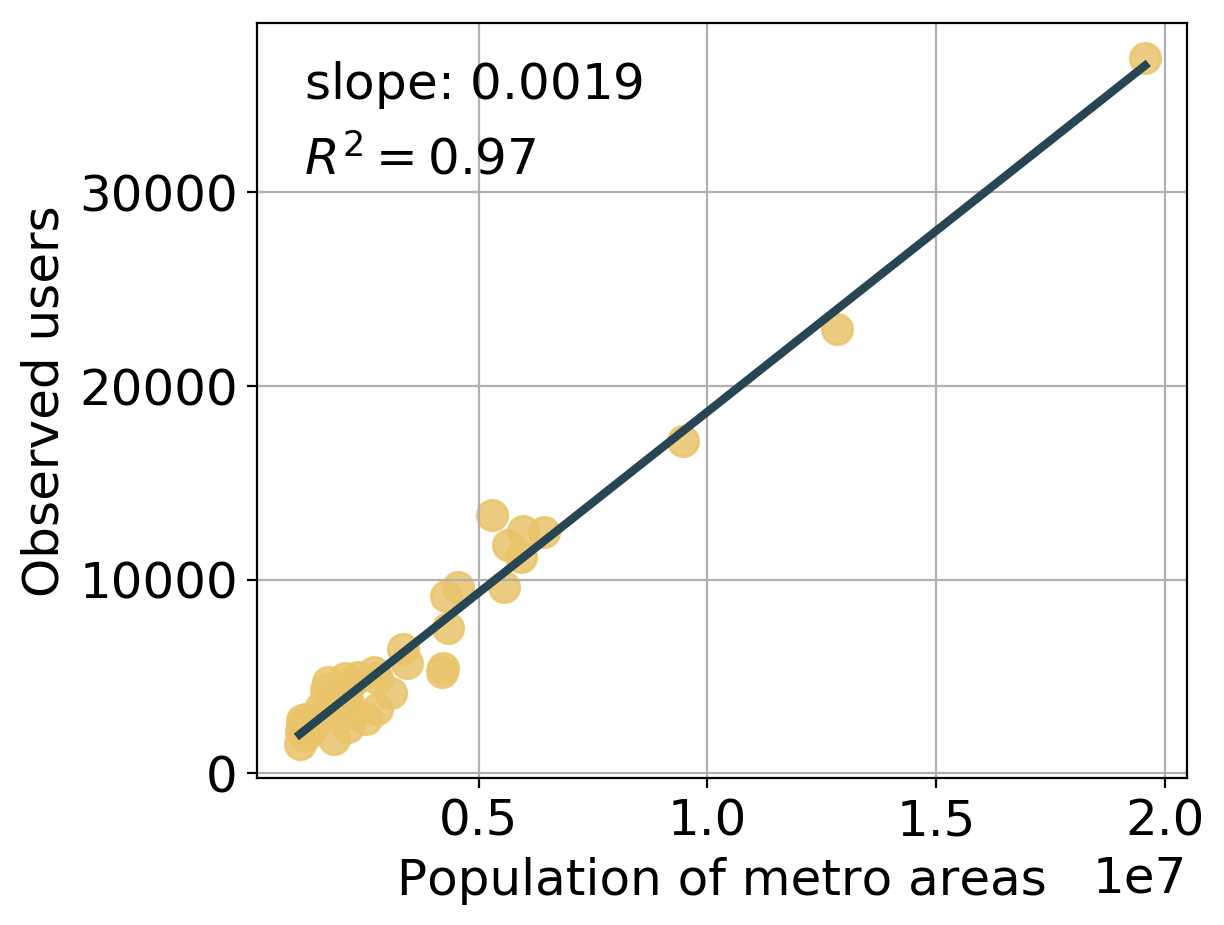}
\caption{\footnotesize 
Population size and observed users in the selected 50 metropolitan areas of the US. Observed users have detected home and work locations, commute at least 100 meter and have at least 1 friendship tie to users with discovered home and work locations inside the same metro area.
}
\end{figure}

\subsection*{SI 3: Distribution of commuting distances}

\begin{figure}[htb]
\centering
\includegraphics[width=145mm]{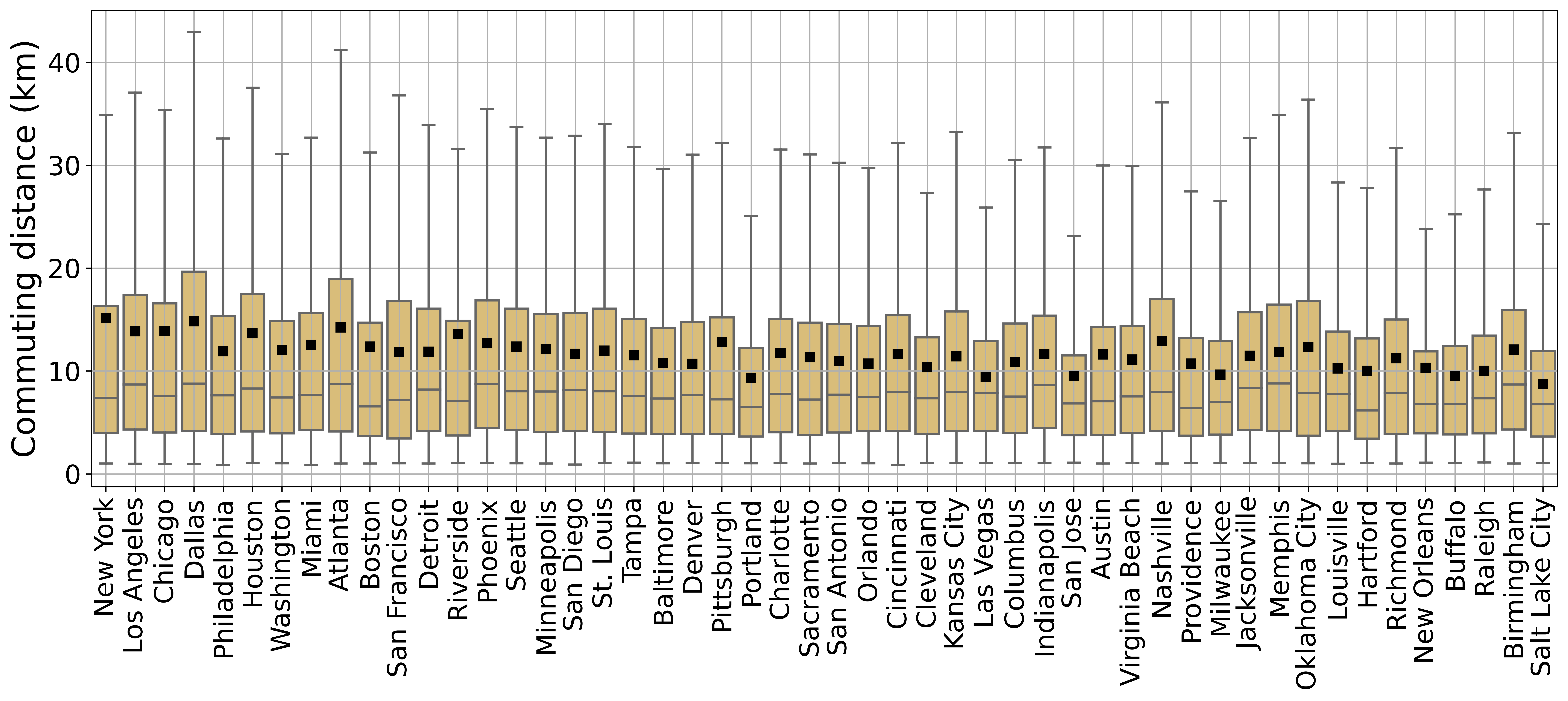}
\caption{\footnotesize 
The distribution of commuting distances in the selected 50 metropolitan areas represented by boxplots. Black dots represent the average commuting distance in each metro area in our data. We only consider those users for whom home and work locations are identifiable, home and work is separated by a minimum 100 meter commute, and the user has minimum 1 friend with identified home and work locations in the same metro area. The metro areas are ordered by population and outlier individuals are not presented. 
}
\end{figure}

\clearpage

\subsection*{SI 4: Effect of different distance thresholds}

\begin{figure}[htb]
\centering
\includegraphics[width=\textwidth]{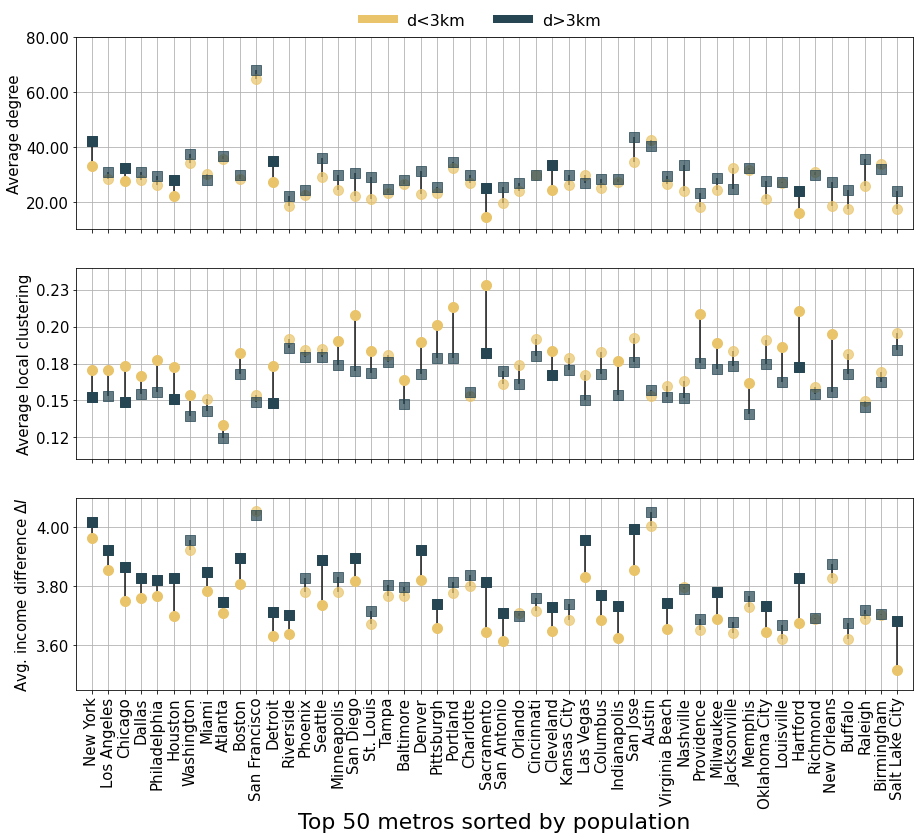}
\caption{\footnotesize
Network characteristics of users commuting above and below 3 km distance in the top 50 metropolitan areas of the United States. The slightly more transparent signs indicate that differences of means are not significant (p$>$0.05).
}
\label{fig:rob3}
\end{figure}

\begin{figure}[htb]
\centering
\includegraphics[width=\textwidth]{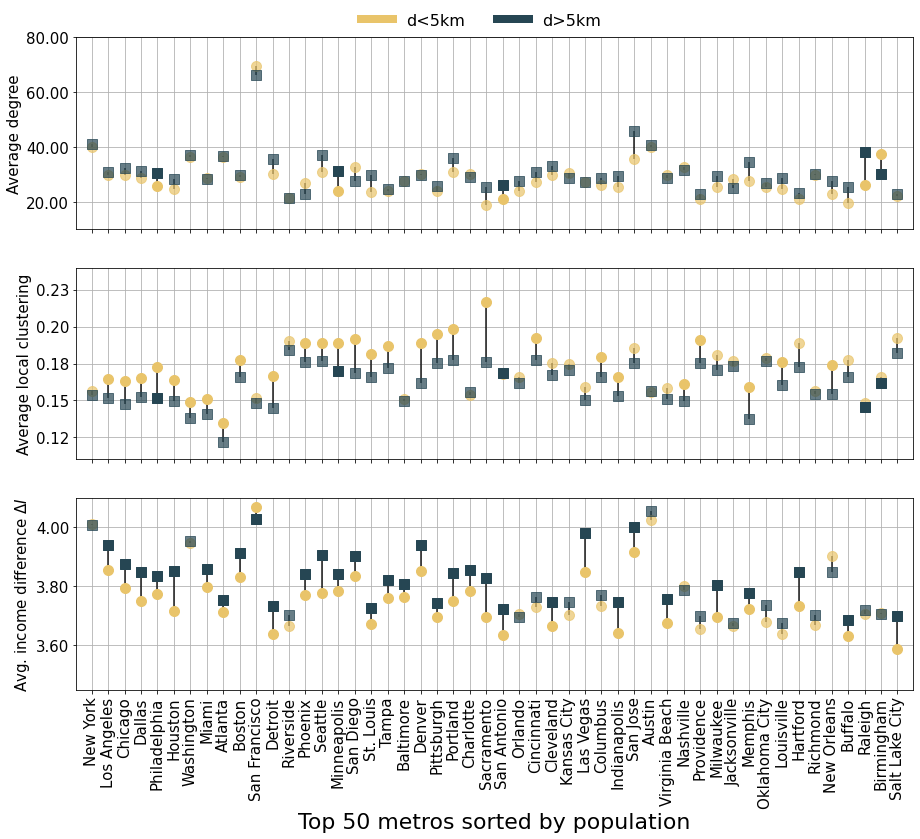}
\caption{\footnotesize
Network characteristics of users commuting above or below 5 km distance in the top 50 metropolitan areas of the United States. The slightly more transparent signs indicate that differences of means are not significant (p$>$0.05).
}
\label{fig:rob5}
\end{figure}

\begin{figure}[htb]
\centering
\includegraphics[width=\textwidth]{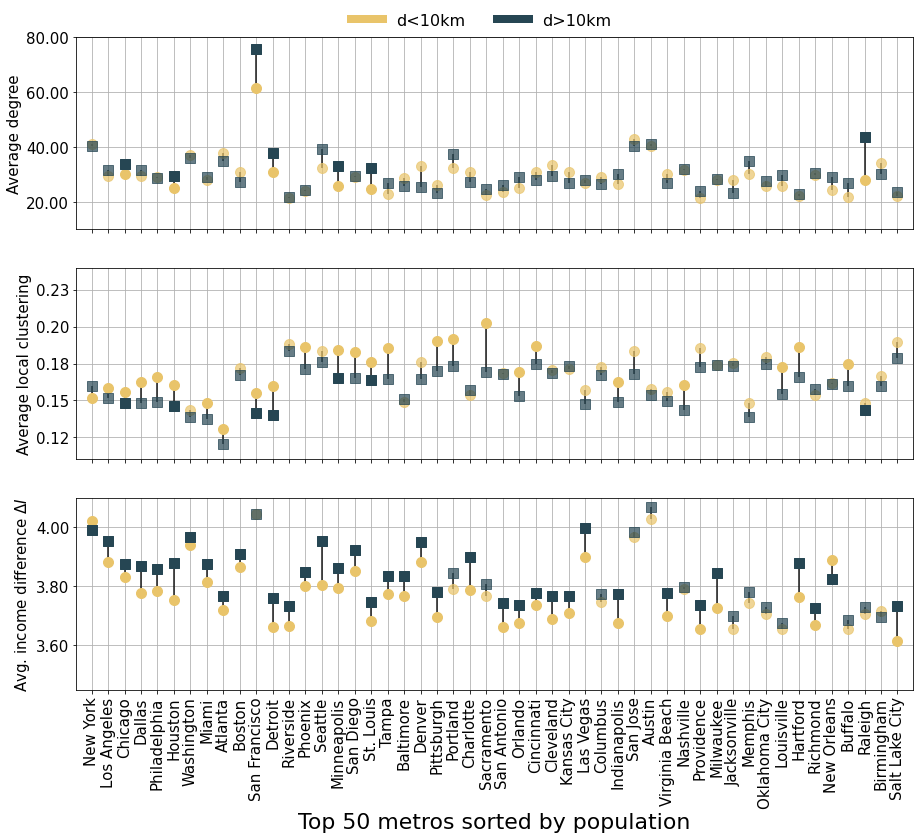}
\caption{\footnotesize
Network characteristics of users commuting above or below 10 km distance in the top 50 metropolitan areas of the United States. The slightly more transparent signs indicate that differences of means are not significant (p$>$0.05).
}
\label{fig:rob10}
\end{figure}

\clearpage

\subsection*{SI 5: Regression on network characteristics and commuting}

\noindent
Table 1 presents 8 linear regression models to complement Figure 2 of the main text. For these robustness checks we log transferred (indicated in the table) or normalized the variables. We introduce control variables step-by-step. Model (1)-(4) further strengthens our previous findings at Figure 2a as longer commuting is connected to lower local clustering in the mutual followership ego network of users. However, the positive and significant quadratic term suggests that commuting distance has an increasing return on network clustering. This relationship is stable even while controlling for the degree, home income and metro area of users. 
Model (5) shows that longer commutes are linked to ego networks with lower income difference between friends and commuting distance has a diminishing return on income difference to friends. However, this relationship does not hold while controlling for the degree and home income of users in Model (8) whereas controllers introduced in Models (6)-(7) are stable.

\begin{table}[!htbp] \centering 
  \caption{Relationship between commuting and network characteristics} 
  \label{} 
\small 
\begin{tabular}{@{\extracolsep{-22pt}}lD{.}{.}{-3} D{.}{.}{-3} D{.}{.}{-3} D{.}{.}{-3} D{.}{.}{-3} D{.}{.}{-3} D{.}{.}{-3} D{.}{.}{-3} } 
\\[-1.8ex]\hline 
\hline \\[-1.8ex] 
 & \multicolumn{8}{c}{\textit{Dependent variable}} \\ 
\cline{2-9} 
\\[-1.8ex] & \multicolumn{4}{c}{Local clustering} & \multicolumn{4}{c}{Income diff. (log)} \\ 
\\[-1.8ex] & \multicolumn{1}{c}{(1)} & \multicolumn{1}{c}{(2)} & \multicolumn{1}{c}{(3)} & \multicolumn{1}{c}{(4)} & \multicolumn{1}{c}{(5)} & \multicolumn{1}{c}{(6)} & \multicolumn{1}{c}{(7)} & \multicolumn{1}{c}{(8)}\\ 
\hline \\[-1.8ex] 
 Distance (log) & -0.056^{***} &  &  & -0.036^{***} & 0.078^{**} &  &  & -0.035 \\ 
  & (0.003) &  &  & (0.003) & (0.031) &  &  & (0.028) \\ 
  & & & & & & & & \\ 
 Distance$^2$ (log) & 0.023^{***} &  &  & 0.013^{***} & -0.059^{***} &  &  & 0.008 \\ 
  & (0.002) &  &  & (0.001) & (0.016) &  &  & (0.014) \\ 
  & & & & & & & & \\ 
 Degree (log) &  & -0.178^{***} &  & -0.179^{***} &  & 0.368^{***} &  & 0.422^{***} \\ 
  &  & (0.001) &  & (0.001) &  & (0.005) &  & (0.005) \\ 
  & & & & & & & & \\ 
 Income (log) &  &  & 0.018^{***} & -0.003^{**} &  &  & 3.506^{***} & 3.546^{***} \\ 
  &  &  & (0.001) & (0.001) &  &  & (0.013) & (0.013) \\ 
  & & & & & & & & \\ 
 Constant & 0.155^{***} & 0.369^{***} & 0.049^{***} & 0.404^{***} & 1.049^{***} & 0.651^{***} & -14.394^{***} & -15.014^{***} \\ 
  & (0.002) & (0.002) & (0.007) & (0.006) & (0.020) & (0.016) & (0.059) & (0.060) \\ 
  & & & & & & & & \\ 
\hline \\[-1.8ex] 
Metro FE & \multicolumn{1}{c}{Yes} & \multicolumn{1}{c}{Yes} & \multicolumn{1}{c}{Yes} & \multicolumn{1}{c}{Yes} & \multicolumn{1}{c}{Yes} & \multicolumn{1}{c}{Yes} & \multicolumn{1}{c}{Yes} & \multicolumn{1}{c}{Yes} \\ 
\hline \\[-1.8ex] 
Observations & \multicolumn{1}{c}{261,283} & \multicolumn{1}{c}{261,283} & \multicolumn{1}{c}{258,949} & \multicolumn{1}{c}{258,949} & \multicolumn{1}{c}{348,728} & \multicolumn{1}{c}{348,728} & \multicolumn{1}{c}{345,610} & \multicolumn{1}{c}{345,610} \\ 
R$^{2}$ & \multicolumn{1}{c}{0.009} & \multicolumn{1}{c}{0.243} & \multicolumn{1}{c}{0.007} & \multicolumn{1}{c}{0.244} & \multicolumn{1}{c}{0.009} & \multicolumn{1}{c}{0.023} & \multicolumn{1}{c}{0.182} & \multicolumn{1}{c}{0.200} \\ 
Adjusted R$^{2}$ & \multicolumn{1}{c}{0.008} & \multicolumn{1}{c}{0.243} & \multicolumn{1}{c}{0.007} & \multicolumn{1}{c}{0.244} & \multicolumn{1}{c}{0.009} & \multicolumn{1}{c}{0.023} & \multicolumn{1}{c}{0.182} & \multicolumn{1}{c}{0.200} \\ 
\hline 
\hline \\[-1.8ex] 
\textit{Note:}  & \multicolumn{8}{r}{$^{*}$p$<$0.1; $^{**}$p$<$0.05; $^{***}$p$<$0.01} \\ 
\end{tabular} 
\end{table}

\clearpage
\noindent
Table 2 presents 3 additional linear regression models to uncover the relationship between the degree of users and their commuting distance. Results are in line with the trends of SI 4 Figure 8-10 as longer commuting is connected to higher degree, however, commuting distance has diminishing returns on user degree.

\begin{table}[!htbp] \centering 
  \caption{Relationship between degree and commuting} 
  \label{} 
\small 
\begin{tabular}{@{\extracolsep{-5pt}}lD{.}{.}{-3} D{.}{.}{-3} D{.}{.}{-3} } 
\\[-1.8ex]\hline 
\hline \\[-1.8ex] 
 & \multicolumn{3}{c}{\textit{Dependent variable}} \\ 
\cline{2-4} 
\\[-1.8ex] & \multicolumn{3}{c}{Degree} \\ 
\\[-1.8ex] & \multicolumn{1}{c}{(1)} & \multicolumn{1}{c}{(2)} & \multicolumn{1}{c}{(3)}\\ 
\hline \\[-1.8ex] 
 Distance (log) & 0.146^{***} &  & 0.144^{***} \\ 
  & (0.010) &  & (0.010) \\ 
  & & & \\ 
 Distance$^2$ (log) & -0.074^{***} &  & -0.073^{***} \\ 
  & (0.005) &  & (0.005) \\ 
  & & & \\ 
 Income (log) &  & -0.096^{***} & -0.096^{***} \\ 
  &  & (0.005) & (0.005) \\ 
  & & & \\ 
 Constant & 1.047^{***} & 1.528^{***} & 1.474^{***} \\ 
  & (0.007) & (0.021) & (0.021) \\ 
  & & & \\ 
\hline \\[-1.8ex] 
Metro FE & \multicolumn{1}{c}{Yes} & \multicolumn{1}{c}{Yes} & \multicolumn{1}{c}{Yes} \\ 
\hline \\[-1.8ex] 
Observations & \multicolumn{1}{c}{348,728} & \multicolumn{1}{c}{345,610} & \multicolumn{1}{c}{345,610} \\ 
R$^{2}$ & \multicolumn{1}{c}{0.007} & \multicolumn{1}{c}{0.007} & \multicolumn{1}{c}{0.008} \\ 
Adjusted R$^{2}$ & \multicolumn{1}{c}{0.007} & \multicolumn{1}{c}{0.007} & \multicolumn{1}{c}{0.008} \\ 
\hline 
\hline \\[-1.8ex] 
\textit{Note:}  & \multicolumn{3}{r}{$^{*}$p$<$0.1; $^{**}$p$<$0.05; $^{***}$p$<$0.01} \\ 
\end{tabular} 
\end{table}

\clearpage

\newgeometry{left=1cm,bottom=2cm,right=1cm,top=2cm}

\subsection*{SI 6: All assortativity matrices for the top 50 US metropolitan areas}

\begin{figure}[h!]
\centering
\includegraphics[width=0.8\textwidth]{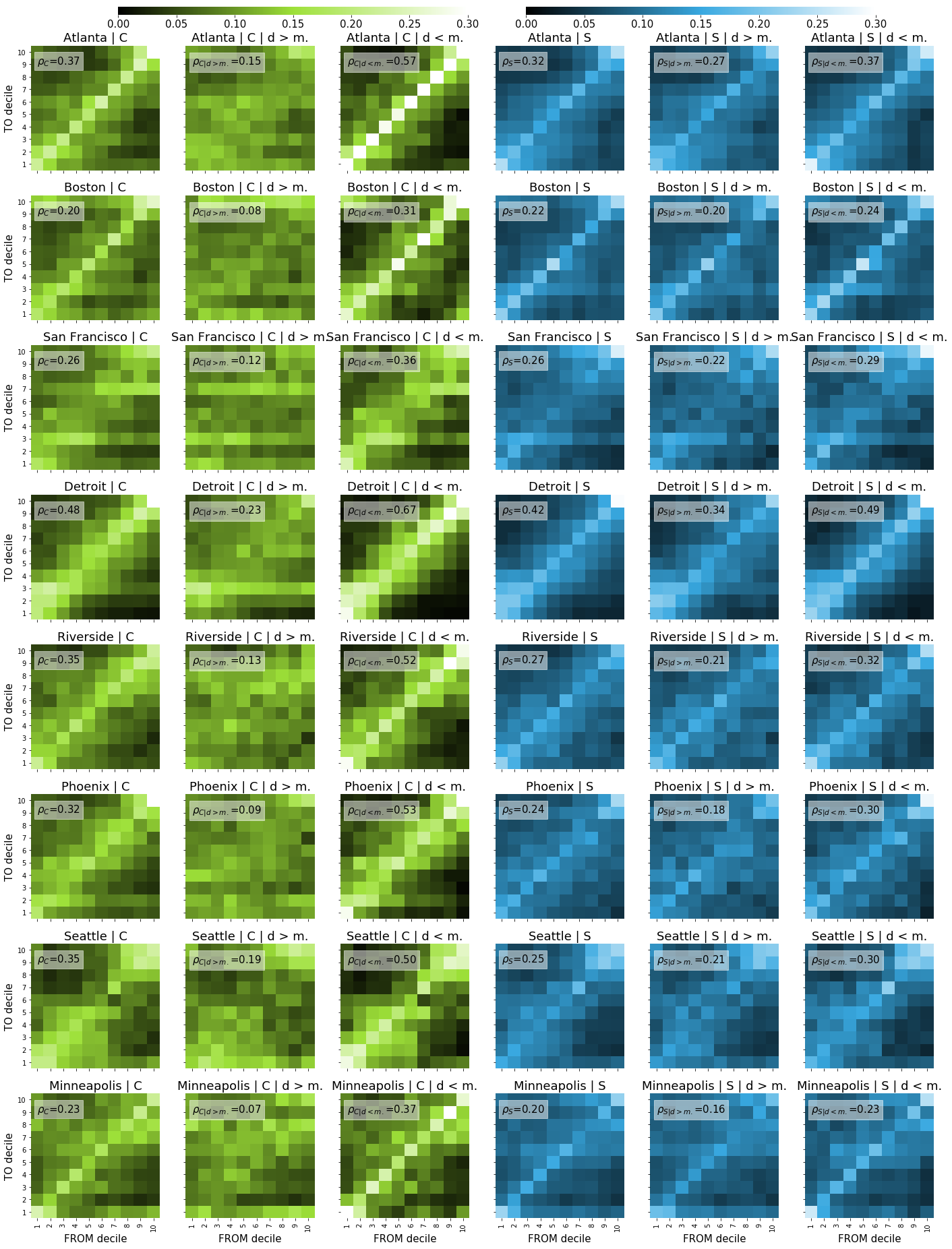}
\caption{\footnotesize Assortativity matrices $C$ and $S$ in all 50 investigated US metropolitan areas for the overall, mobile and non-mobile users. $\rho$-values are indicated in the labels.}
\label{fig:50matrixC}
\end{figure}

\clearpage

\setcounter{figure}{10}

\begin{figure}[htb]
\centering
\includegraphics[width=0.8\textwidth]{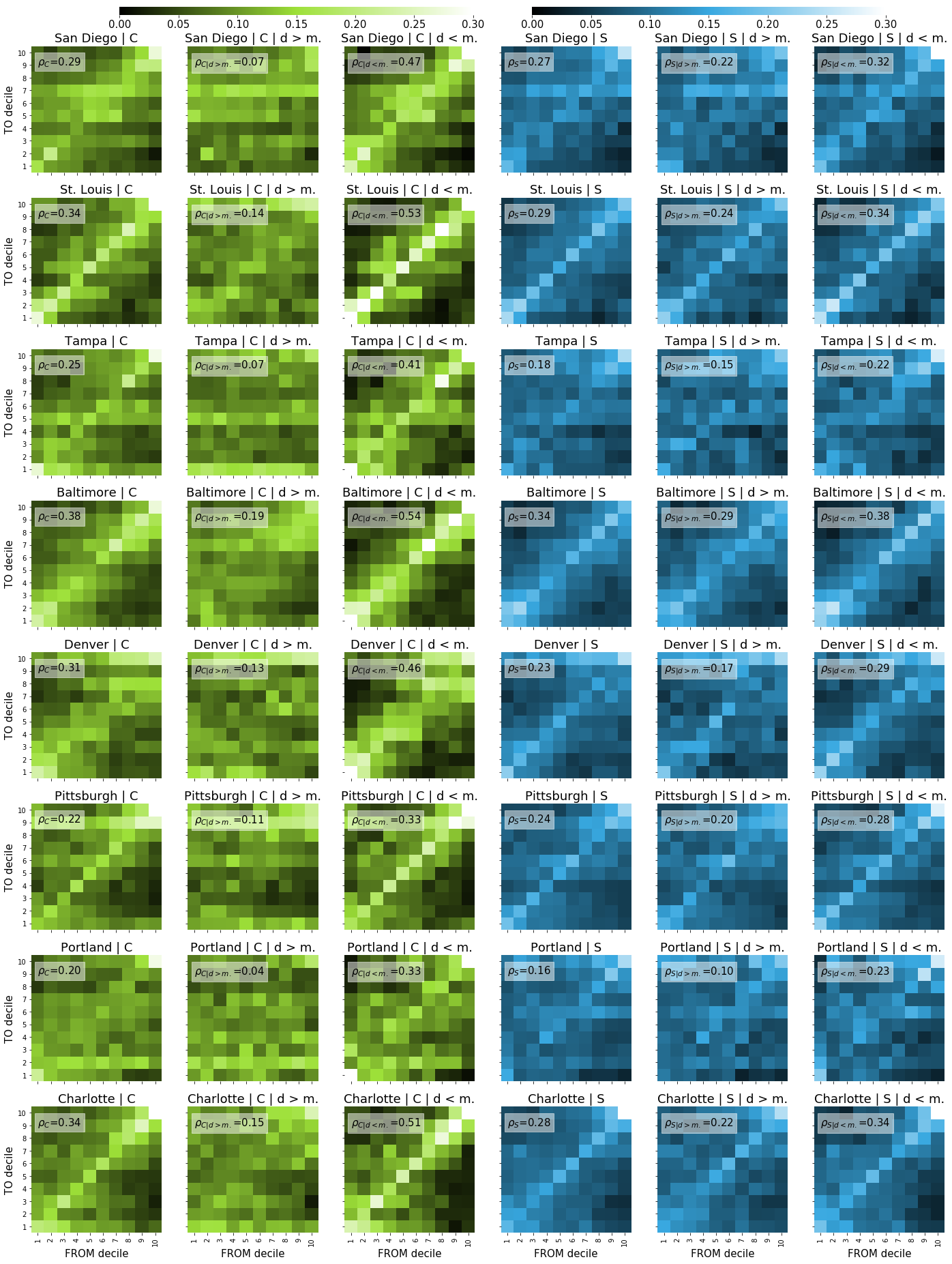}
\caption{\footnotesize (continued) Assortativity matrices $C$ and $S$ in all 50 investigated US metropolitan areas for the overall, mobile and non-mobile users. $\rho$-values are indicated in the labels.}
\end{figure}

\clearpage

\setcounter{figure}{10}

\begin{figure}[htb]
\centering
\includegraphics[width=0.8\textwidth]{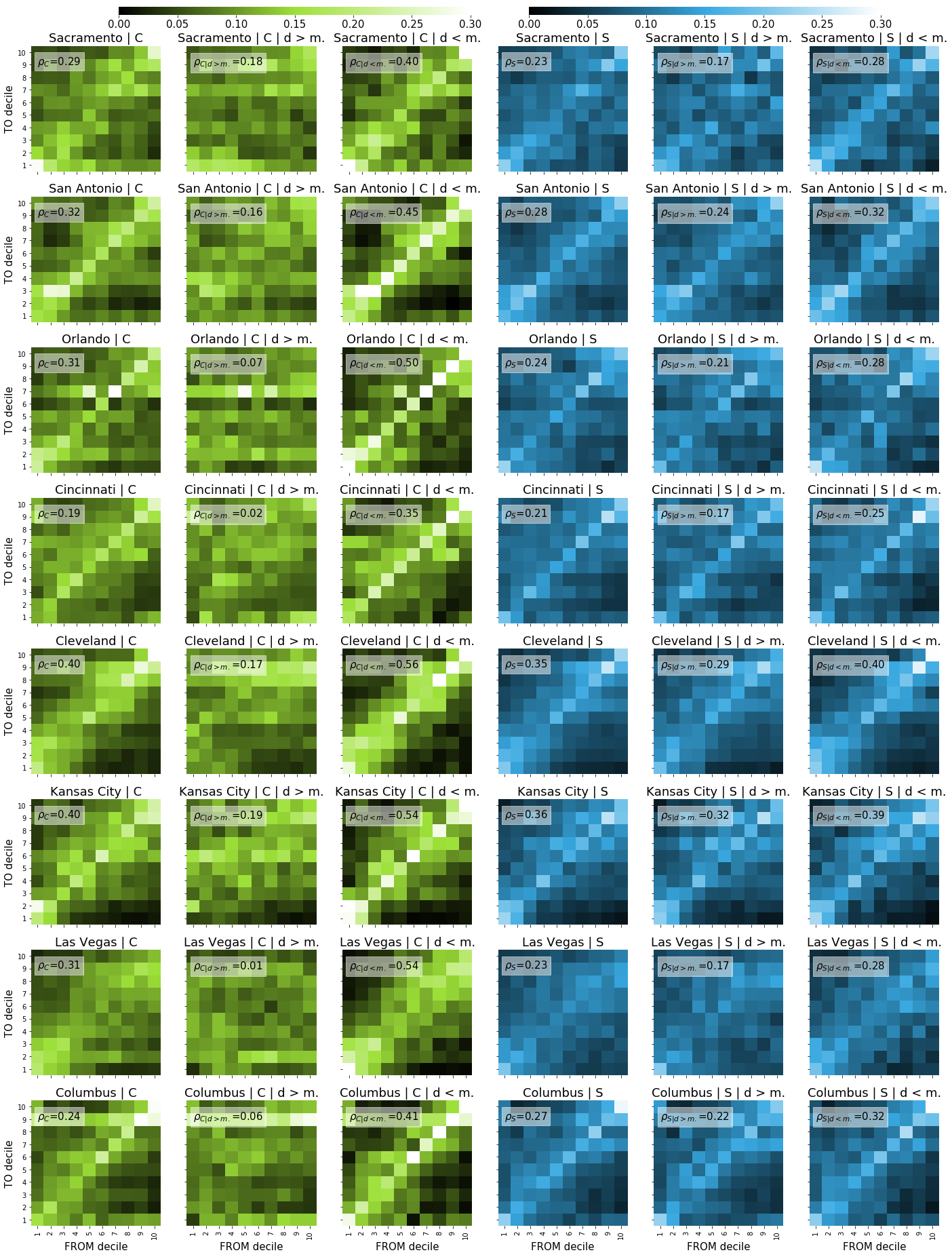}
\caption{\footnotesize (continued) Assortativity matrices $C$ and $S$ in all 50 investigated US metropolitan areas for the overall, mobile and non-mobile users. $\rho$-values are indicated in the labels.}
\end{figure}

\clearpage

\setcounter{figure}{10}

\begin{figure}[htb]
\centering
\includegraphics[width=0.8\textwidth]{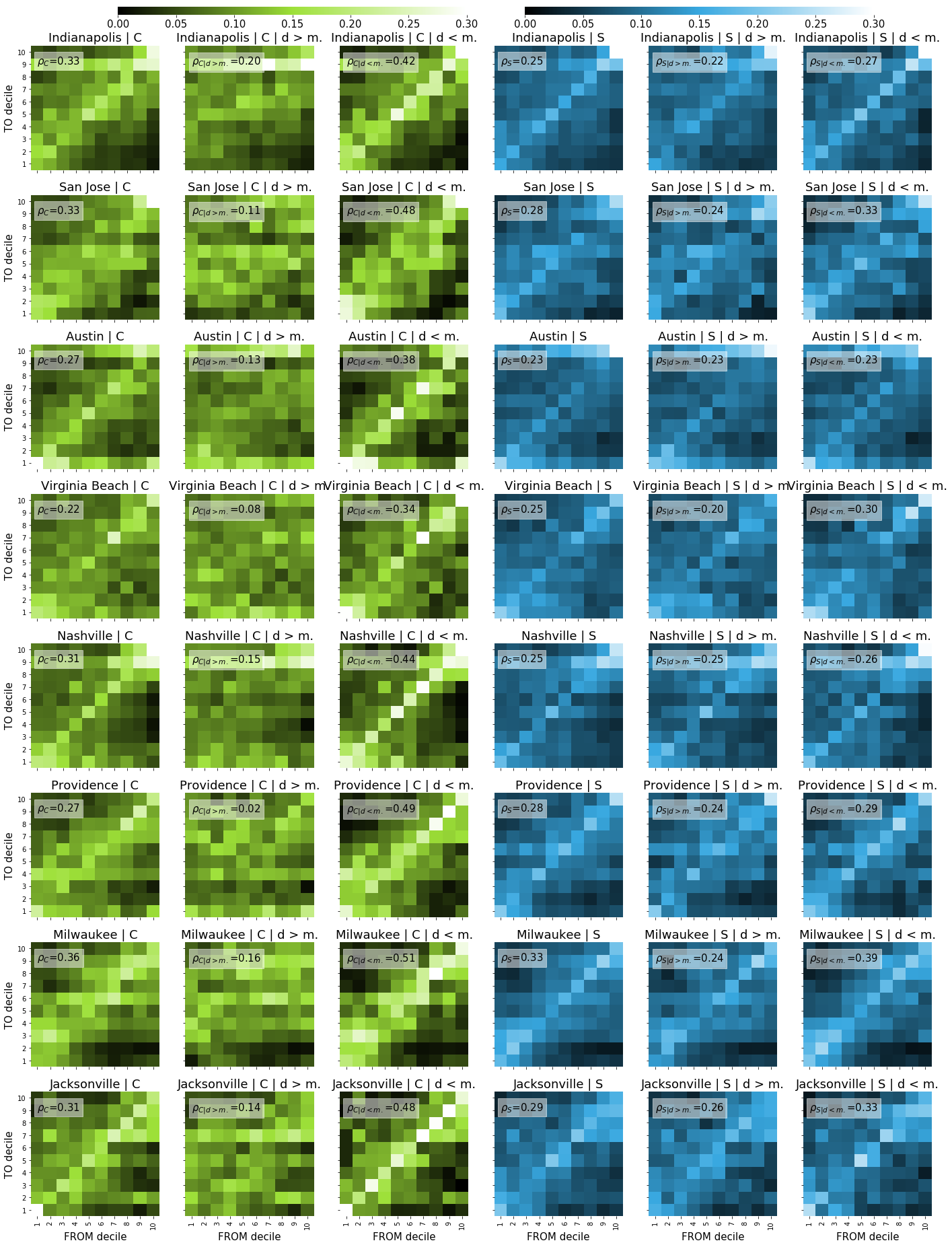}
\caption{\footnotesize (continued) Assortativity matrices $C$ and $S$ in all 50 investigated US metropolitan areas for the overall, mobile and non-mobile users. $\rho$-values are indicated in the labels.}
\end{figure}

\clearpage

\setcounter{figure}{10}

\begin{figure}[htb]
\centering
\includegraphics[width=0.8\textwidth]{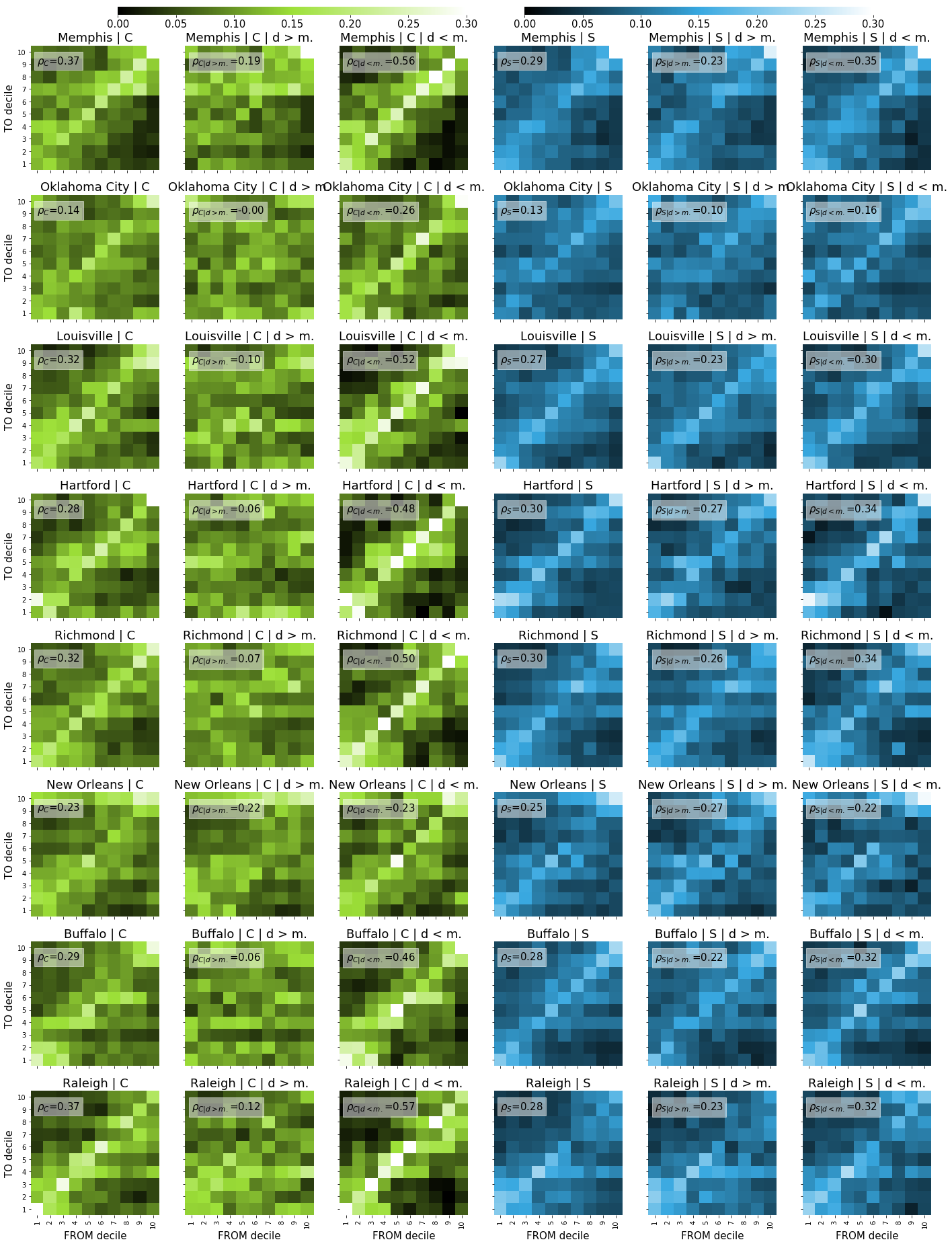}
\caption{\footnotesize (continued) Assortativity matrices $C$ and $S$ in all 50 investigated US metropolitan areas for the overall, mobile and non-mobile users. $\rho$-values are indicated in the labels.}
\end{figure}

\clearpage

\setcounter{figure}{10}

\begin{figure}[htb]
\centering
\includegraphics[width=0.8\textwidth]{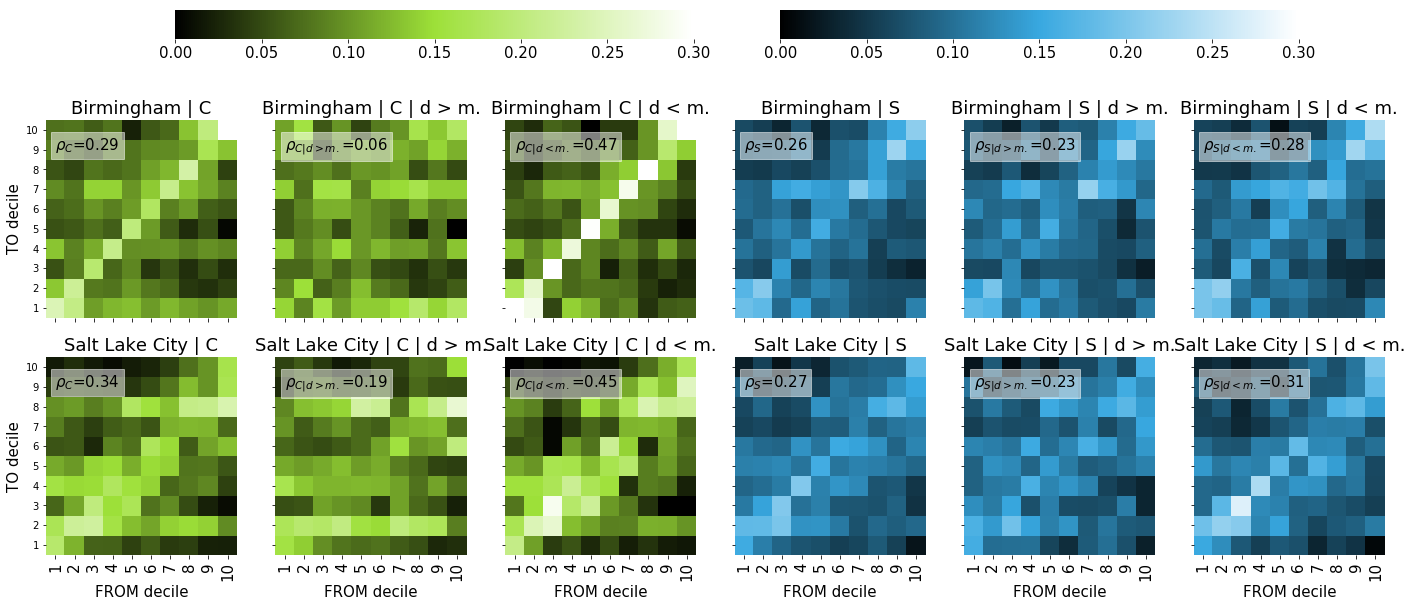}
\caption{\footnotesize (continued) Assortativity matrices $C$ and $S$ in all 50 investigated US metropolitan areas for the overall, mobile and non-mobile users. $\rho$-values are indicated in the labels.}
\end{figure}

\clearpage

\restoregeometry

\subsection*{SI 7: Different distance thresholds and assortativity change}

\begin{figure}[htbp!]
    \centering
    \includegraphics[width=\textwidth]{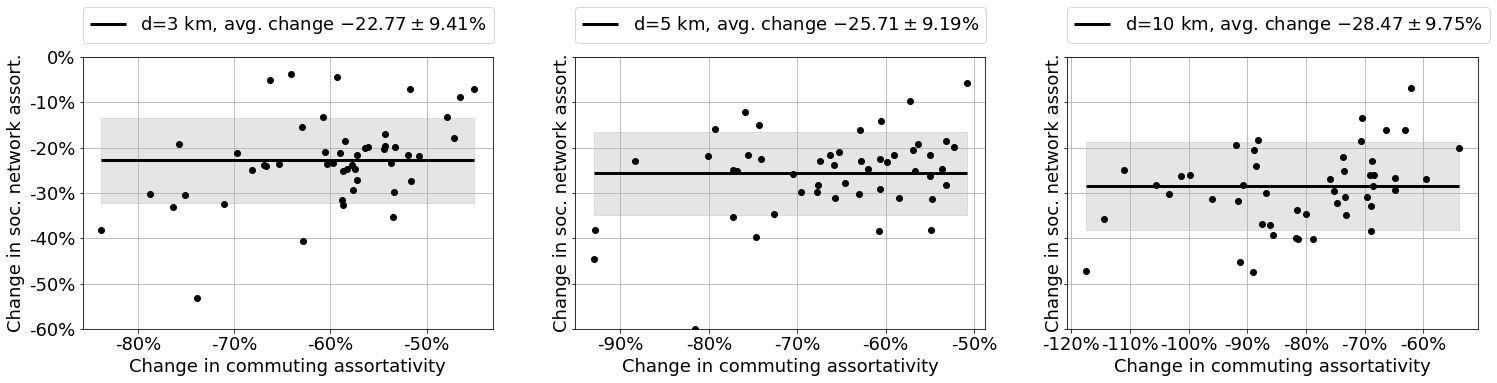}
    \caption{Change in the assortativity of the social network matrices vs. the commuting matrices for different distance thresholds}
    \label{fig:changes}
\end{figure}

\begin{figure}[htbp!]
    \centering
    \includegraphics[width=10cm]{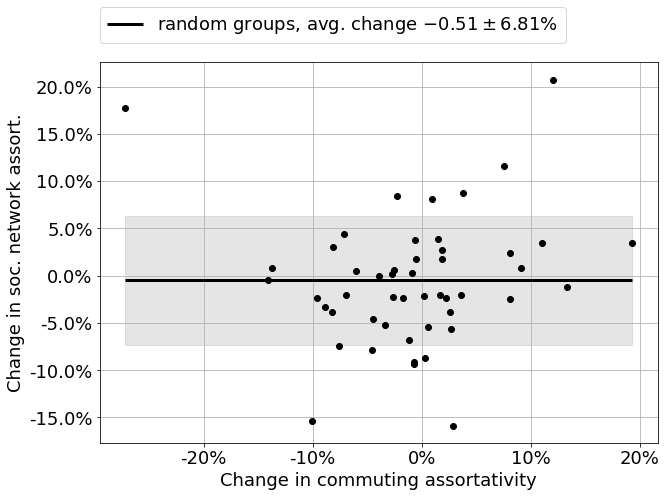}
    \caption{Change in the assortativity of the social network matrices vs. the commuting matrices for two random user groups}
    \label{fig:random}
\end{figure}

\clearpage

\subsection*{SI 8: Diversity in commuting and social connections}

We measure the diversity $S_C$ and $S_S$ of the matrices $C$ and $S$ (or for any matrices on the smaller user base, e.g. $S_{d>{median}}$) by averaging the normalized entropies of the columns of the matrices. Formally, for a  $10\times 10$ matrix $X$, where the sum of the columns $\sum_j X_{ij}=1$ for every possible $j$,
\begin{equation}
    S_X = \frac{1}{10}\cdot \sum\limits_{j=1}^{10}\frac{1}{\log 10}\sum\limits_{i=1}^{10}X_{ij}\cdot\log X_{ij},
    \label{eq:diversity}
\end{equation}
which means that $S_X=0$ corresponds to a matrix in which every column contains exactly one element that is 1, and the others are 0, and $S_X=1$ corresponds to the case when every element of the matrix is equal, $\frac{1}{10}$. Thus, $S_X$ values closer to 1 mean matrices in which commuting or friendship ties in a column are on average more distributed over multiple income classes, whereas smaller $S_X$ values mean matrix columns with rather one dominant element.

In parallel to the decreasing assortativity with longer commutes, we can observe an increasing average diversity for the connection patterns of both matrices, if measured by the averaged entropy of the column-wise probability distributions $S_S$ and $S_C$ (see Section~\ref{sec:mm} for details on this measure). This increase in the diversity is shown for all 50 metropolitan areas. Again, there is a higher increase in diversity for the commuting assortativity matrix, if we compare long commuters to short commuters, but this increase in the diversity is in parallel with the increase in the friendship assortativity matrix. If we measure which income deciled contribute to the increasing entropy values in both the mobility and the social network patterns, we can see that the lowest and highest income classes have the most diversity increase (see the inset). Therefore, it is most likely that rearrangement of the social connections of the richest and poorest deciles contribute most to the 30\% decrease in social network assortativity that comes with longer commutes.

\begin{figure}[htbp!]
    \centering
    \includegraphics[width=0.9\textwidth]{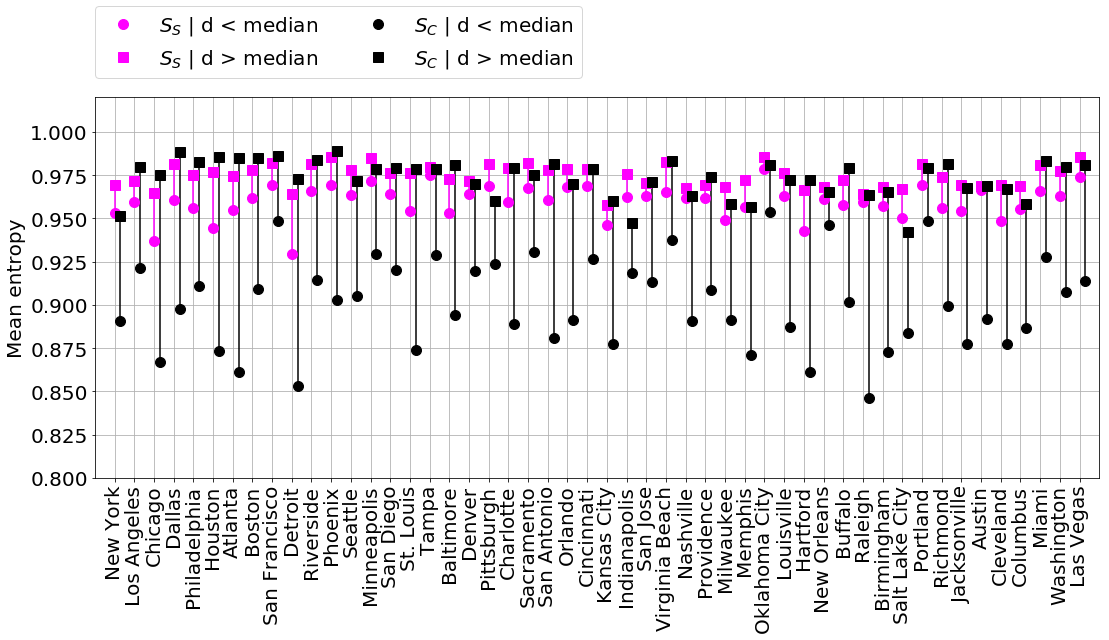}
    \caption{Diversity}
    \label{fig:diversity}
\end{figure}

\begin{figure}[htbp!]
    \centering
    \includegraphics[width=0.9\textwidth]{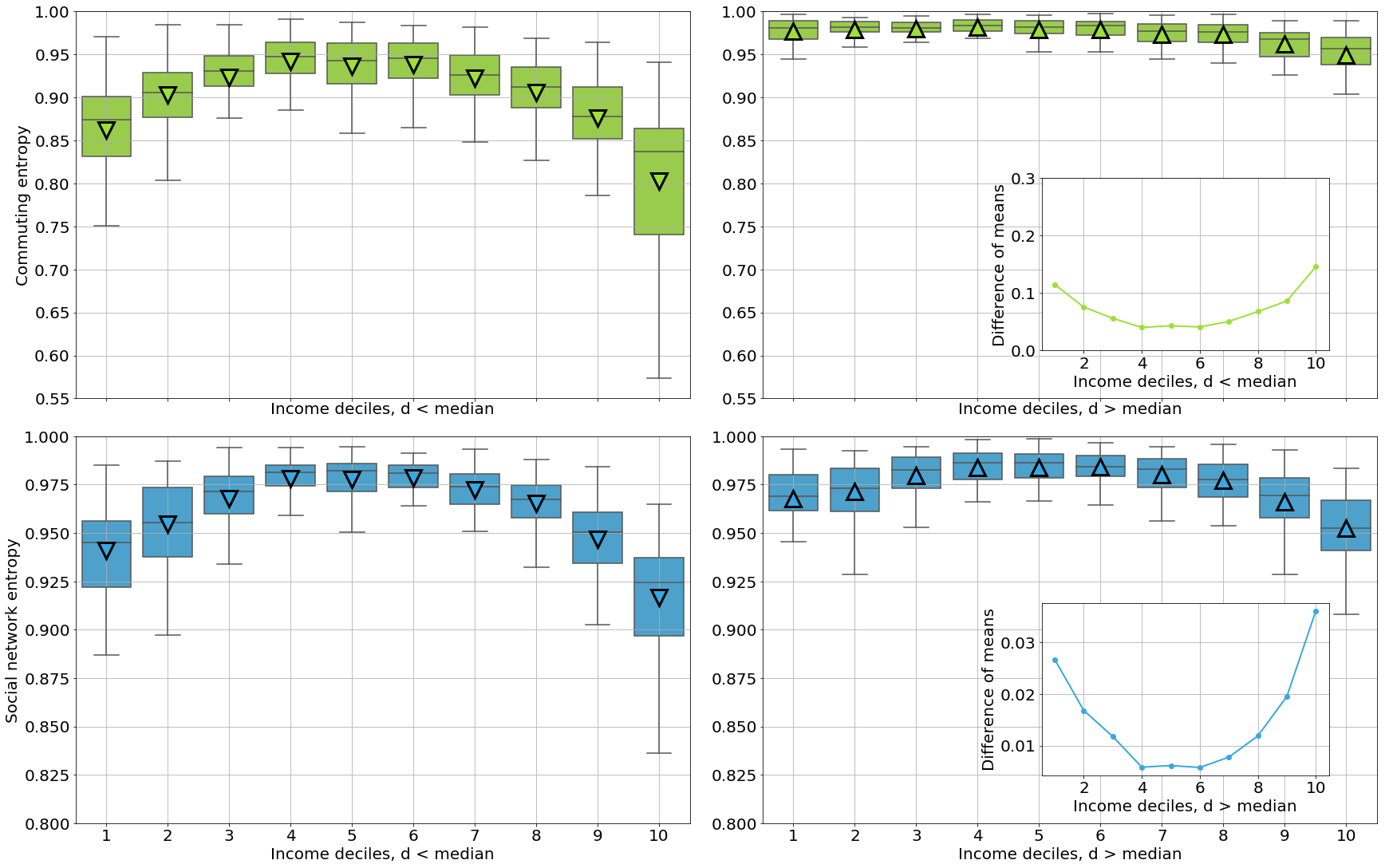}
    \caption{Assortativity of commuting and social network matrices by income groups.}
    \label{fig:entropy}
\end{figure}

\end{document}